\documentclass[10pt,twocolumn,5p,sort&compress]{elsarticle}
\usepackage{amssymb}
\usepackage{color}
\newcommand{\be}{\begin{equation}}
\newcommand{\ee}{\end{equation}}
\newcommand{\bea}{\begin{eqnarray}}
\newcommand{\eea}{\end{eqnarray}}
\begin{document}
\begin{frontmatter}
 \title{
The generalized Langevin equation revisited: Analytical expressions for the persistence dynamics of a viscous fluid under a time dependent external force.}
\author[wor]{Wilmer Olivares--Rivas}
\ead{wilmer@ula.ve}
\author[pjc]{Pedro J. Colmenares\corref{cor1}}
\ead{colmenar@ula.ve}
\cortext[cor1]{Corresponding author}
\address{Grupo de Qu\'{\i}mica Te\'orica: Quimicof\'{\i}sica de Fluidos y Fen\'omenos Interfaciales (QUIFFIS).
Departamento de Qu\'{\i}mica -- Universidad de Los Andes. M\'erida 5101, Venezuela}
\begin{abstract}
The non--static generalized Langevin equation and its corresponding
Fokker--Planck equation for the position of a viscous fluid particle were solved in closed form for a time dependent external force. Its solution for a constant external force was obtained analytically. The non--Markovian stochastic differential
equation, associated to the dynamics of the position under a colored noise, was then applied to the description of the dynamics and persistence time of particles constrained within absorbing barriers. Comparisons with molecular dynamics were very satisfactory.
\end{abstract}
\begin{keyword}
Stochastic processes, Langevin method, diffusion constant, molecular dynamics.

Published in Physica A 458 (2016) 76-94

doi: 10.1016/j.physa.2016.03.112
\end{keyword}
\end{frontmatter}


\section{Introduction.}
\label{intro}
The dynamics of particles in a potential field is usually investigated by molecular dynamics (MD). However, since
the time scale of such a procedure is of the order of femtoseconds, the algorithms are in general very expensive
in computer consuming time. To overcome this difficulty, one generally appeals to mesoscopic descriptions
on which the time scale is larger. It gives a glimpse about the dynamical behavior of the system at the time scale of the technique. There are three levels of description. The most commonly used description is the so called Langevin equation (LE), where the dynamics of a Brownian particle in phase space is described
by the Markovian set of stochastic differential equations (SDE)\cite{mcquarrie,joel,reichl}
\bea
m\frac{d\,v(z,t)}{dt}\!&=& -\gamma(z)\, v(z,t)+\eta(z)\,\xi(t)\!+ F(z,t),\nonumber\\
\dot{z}&=&v,
\label{setLE}
\eea
where $z(t)$ and $v(z,t)$ are the position and velocity of the tagged particle of mass $m$ at time $t$, respectively; $\gamma(z)$ is the
position--dependent phenomenological fluid friction
coefficient; $F(z,t)$ is the
external force field, $\eta(z)$ is the intensity of the stochastic force and $\xi(t)$ is a zero--mean Gaussian white noise term. The static LE can be solved by a double
integration or alternatively, through the solution of a second order differential equation\cite{mcquarrie,reichl}. In fact, an analytical solution for a confined fluid has been previously given\cite{CLOR}.

An important simplification of the LE with a wide range of practical applications is the so-called Smoluchowski limit also known as the high friction limit (HFL). The most simple derivation consists on assuming that the fluid has a large enough
friction coefficient $\gamma$ so that the velocity
in the LE shows a fast relaxation to a quasi equilibrium
\cite{gardiner}, with $dv/dt=0$.
Thus, position evolves with time according to
\be
\frac{d z}{dt}=v_{_{\mathrm{F}}}(z) + \sqrt{2D_0}\,\xi(t) ,
\ee
where $v_{_{\mathrm{F}}}(z)=F(z,t)/\gamma(z)$
is the induced velocity drift and
$D_0=(1/2)(\eta(z)/\gamma(z))^{2}$
 is interpreted as a diffusion coefficient. The corresponding
master equation for the probability density corresponding to this SDE is the Smoluchowski equation (SE). Clearly this is only valid in the long time limit, also
known as the Smoluchowski regime.
The third stochastic description is an improvement without restriction on the time scale, known as the non-static or \emph{generalized Langevin equation (GLE)}. It is obtained modifying the dissipation or friction term
\bea
m\frac{d\,v(z,t)}{dt}&=&-m\int_{0}^{t}\Gamma(z,t-t^{\prime}) \,v(z,t^{\prime})\,dt^{\prime}\nonumber\\
&&+ R(t) + F(z,t),
\label{GLE}
\eea
where the kernel $\Gamma(z,t-t^{\prime})$ expresses the memory or retardation effect
on the movement of the fluid
particle due to the collective hydrodynamic response of the surrounding fluid \cite{boom} and, $R(t)=\eta(z,t)J(t)$ a
colored--Gaussian fluctuating driving force. Now the random noise function $J(t)$ is not a white noise. If the retardation is omitted, the kernel is static
$\Gamma(z,t-t^{\prime})=\alpha(z)\,\delta(t-t^{\prime})$ with $\alpha(z)=\gamma(z)/m$, and the LE, Eq. (\ref{setLE}), is
recovered. This is equivalent to take the noise as being white, namely, $J(t)=\xi(t)$.
However, when inertial effects are not negligible -the friction coefficient is not dominant in the dynamics-- the position and velocity are driven by non-markovian random terms having a finite correlation time. This is the situation for real non homogeneous fluids and systems with strong boundary conditions or in the presence of chemical reactions.

A well known work obtaining a master equation for the probability density $p(\zeta,t\mid \zeta_{0})$ for a general $\zeta$ process, as a Kramers-Moyal\cite{risken} cumulant expansion, is due to H\"anggi in 1978\cite{hanggi,hanggiTalkner}. However, \textcolor{black}{the GLE, Eq. (\ref{GLE})}, was discussed for a viscous fluid by Chow and Hermans\cite{chow}  \textcolor{black}{as far back as} 1972 and a master equation \textcolor{black}{for that generalized version}, was nicely derived and solved by Dufty\cite{dufty} in 1974, being commonly referred to as the Fokker-Planck equation. In the absence of external forces it was extensively studied by Adelman\cite{adelman1}, Fox\cite{fox1}, Volkov and Pokrovsky\cite{volkov} and later by Rodriguez--Salinas \cite{rodrisalinas}. Numerical approaches for general first--order SDEs with colored random forces have been also presented\cite{budini,darvea}. Budini and C\'aceres\cite{budini}, numerically obtained the velocity distribution associated to the GLE SDE for arbitrary noise and memory kernels, but with no external force. They found that the interplay of noise structure and dissipation is an important issue to consider, in order to achieve the stationary steady state of the probability density.

    In this paper we analyze the dynamical or generalized Langevin approach, Eq.(\ref{GLE}), for the motion of an interacting fluid particle in an external time-dependent field. We obtain the corresponding generalized Fokker-Planck equations (GFPE) for the Bayesian probabilities $p(v,t|v_{0},0)$ and $p(z,t|z_{0},v_{0},0)$. Then, following Chandrasekhar's approach\cite{chandra}, we obtain the $v_{0}$ averaged probability densities $p(v,t)$ and $p(z,t|z_{0},0)$ giving analytical formulas for their moments. We shall discuss the fact that this equation can be written as a much simpler statistically equivalent Markovian first--order SDE. Revisiting the master equation associated to GLE, we get consistent analytical results for the survival probability in a viscous fluid and evaluate important dynamical properties as the mean square displacement (MSD) for a fluid bounded by absorbing barriers.

Analytical expressions for Chandrasekhar conditional probability density $p(z,t|z_{0},v_{0})$ and the corresponding Rayleigh type GFPE in the presence of an external force are presented in Section \ref{sec:I Generalized}.
That is, with the method based on the solution of the stochastic Liouville equation\cite{kubo,vankampen2,sancho}, we find a well defined GRFPE, whose analytical solution matches Chandrasekhar's lemma\cite{chandra}.

    Section \ref{sec:II GFPE} deals with the GLE's velocity and position space distribution moments for any given external time dependent force $F(t)$. First, we revise the GLE $z$-space to find the moments of the probability density $p(z,t|z_{0})$ averaged over the initial velocity, unifying all the equivalent master equations and SDE. The known results on the velocity space probability \cite{adelman1,fox1} are complemented to include a time dependent external force.

Considering an exponential decaying friction coefficient kernel and an external constant force, we derive in section \ref{sec:III Analytical}, analytical results for the moments of the distributions in $z$ and $v$ spaces.
The application of the GLE theory to the problem of survival probability and first passage time
of a constrained fluid is presented in section \ref{sec:IV Dynamical}.

Finally, we include two appendices to show an unified view of the velocity--fluctuation coefficient in the GLE approach and the classical LE and SE results.

\section{The position generalized Rayleigh-Fokker-Planck equation (z-GRFPE).}
\label{sec:I Generalized}
In this section, we use the simple procedures well discussed by Adelman\cite{adelman1}, Fox\cite{fox1} and
Sancho {\it et al.}\cite{sancho}, to obtain the GFPE type master equations associated to the position of the particles of a viscous fluid satisfying the GLE.
 It has been
shown \cite{liu,CLOR} that even for strongly non homogeneous fluid the application of molecular dynamics
and LE equation can be carried out by sampling the z space in virtual layers where the local mean force is
assumed piece-wise constant. Thus, we shall assume that the time dependent force is a constant as a function of position, namely $F(z,t)=F(t)$. Besides the fact that even the LE has no solution for a general $F(z,t)$, it will greatly reduce the mathematical complexity of the equations.

Following the standard method used by Fox\cite{fox1}, we apply the k--Laplace transformation to both
sides of the GLE differential equation, Eq. (\ref{GLE}), and find after collecting terms, an expression for the
Laplace transform of the velocity $\widehat{v}(k)$
\be
\widehat{v}(k)=\widehat{\chi}_{v}(k)\Bigg[v_{_{0}}
+\frac{1}{m}\Big( \widehat{R}(k) + \widehat{F}(k) \Big)\Bigg],
\label{zlaplace}
\ee
 where $v_{_{\!0}}$ is the initial velocity. The function $\widehat{\chi}_{v}(k)$ is
\be
\widehat{\chi}_{v}(k)=\frac{1}{\,k+\widehat{\Gamma}(k)},
\label{chihat}\ee
 with $\widehat{\Gamma}(k)$, $\widehat{R}(k)$, and $\widehat{F}(k)$, being the Laplace transforms of the memory kernel, the colored--noise internal force, and the external force, respectively.

\textcolor{black}{The fundamental Green function $\chi_{v}(t)$ is the inverse transform of $\widehat{\chi}_{v}(k)$.
From the definition in Eq. (\ref{chihat}) we have the
useful Volterra relationship
\be
\dot{\chi}_{v}(t)=\frac{d\,\chi_{v}(t)}{dt}=-\int_{0}^{t}\chi_{v}(s)\,\Gamma(t-s)\,ds.
\label{chivdot}
\ee}
\textcolor{black}{
Before solving Eq. (\ref{zlaplace}), it is useful to define two related functions. So, the integral of $\chi_{v}(t)$ is denoted as $\chi_{z}(t)$ and in turn, the integral of $\chi_{z}(t)$ is denoted simply as $\chi(t)$, without any subscript
\bea
\chi_{z}(t)&=&\int_{0}^{t}\chi_{v}(s)\,ds,\label{chiz1}\\
\chi(t)&=&\int_{0}^{t}\chi_{z}(s)\,ds\label{chiF1}\\
&=& \int_{0}^{t}(t-s)\chi_{v}(s)\,ds.\label{chiFchiv}
\eea
Clearly $\dot{\chi}_{z}=\chi_{v}$ and $\dot{\chi}=\chi_{z}$, with initial conditions
$\chi_{v}(0)=1$, $\chi_{z}(0)=0$, and $\chi(0)=0$.}

\textcolor{black}{
Now, inverting both sides of Eq. (\ref{zlaplace}), we get the particle velocity $v(t)=\dot{z}(t)$ and by integration, its position $z(t)$
\bea
v(t)&=& \overline{v}(t) +\varphi_{v}(t),
\label{vGLE}\\
z(t)&=& \overline{z}(t) + \varphi_{z}(t).
\label{zGLE}
\eea}
\textcolor{black}{The colored noise force response functions $\varphi_{v}(t)$ and $\varphi_{z}(t)$ in the velocity and position components,  are defined respectively as:
\bea
\varphi_{v}(t)&=&\frac{1}{m}\int_{0}^{t}\chi_{v}(t-s)\,R(s)\,ds,
\label{varphi}\\
\varphi_{z}(t)&=&\frac{1}{m}\int_{0}^{t}\chi_{z}(t-s)\,R(s)\,ds.
\label{varphiz}
\eea}
\textcolor{black}{The drift components of the velocity and position are actually the averages over the noise
distribution, namely, $\overline{v}(t)=\langle v(t)\rangle_{_{\!\!R}}$ and
$\overline{z}(t)=\langle z(t)\rangle_{_{\!\!R}}$. They are given as
\bea
\overline{v}(t) &=&v_{_{\!0}}\,\chi_{v}(t)+
\phi_{v}(t),\label{vtilde}\\
\overline{z}(t)&=&z_{0}+v_{_{\!0}} \,\chi_{z}(t)
+\phi_{z}(t).
\label{ztilde}
\eea}
\textcolor{black}{Here $v_{0}$ and $z_{0}$ are the initial velocity and position at $t=0$, and the
corresponding velocity and position response functions to the external force, $ \phi_{v}(t)$ and  $\phi_{z}(t)$, are given by}
 \textcolor{black}{
\bea
\phi_{v}(t)&=&\frac{1}{m}\int_{0}^{t}\chi_{v}(t-s)\,F(s)\,ds,
\label{varphiF}\\
\phi_{z}(t)&=&\frac{1}{m}\int_{0}^{t}\chi_{z}(t-s)\,F(s)\,ds\label{phiF1}\\
&=&\int_{0}^{t}\phi_{v}(s)\,ds.\nonumber
\eea
From the solution for the velocity, Eqs. (\ref{vGLE}), (\ref{varphi} and (\ref{vtilde}), it can be shown that the susceptibility  $\chi_{v}(t)$ is directly related to the \emph{velocity auto-correlation function} (VAC), $c_{v}(t)=\langle v(0)v(t)\rangle\,_{_{\!\!R}}$, namely, $\chi_{v}(t)=c_{v}(t)/c_{v}(0)$.
The noise and external force response functions also have the simple relationships
 $\dot{\varphi_{z}}(t)=\varphi_{v}(t)$ and $\dot{\phi_{z}}(t)=\phi_{v}(t)$. They are functionals of the susceptibility function
 $\chi_{v}(t)$, namely $\varphi_{v}(t)=\varphi_{v}[\chi_{v}(t)]$. Consequently, the conditional probability and the position itself are functionals of $\chi_{v}(t)$, i.e., $p(z,t\!\!\mid\!\! z_{_{0}},v_{0},0)=p[\chi_{v}(t)]$ and $z(t)=z[\chi_{v}(t)]$.}

 \textcolor{black}{The mathematical and physical consistency of the problem requires to know the statistical properties
of $\varphi_{v}(t)$ and $\varphi_{z}(t)$. Above all, they should have zero mean averages, that is
$\langle \varphi_{v}(t)\rangle_{\!R}=0$ and therefore $\langle \varphi_{z}(t)\rangle_{\!R}=0$. They are
described in appendix A, where in particular, we derive the appropriate fluctuation-dissipation theorem.}

 The velocity space has been amply studied. So, in this work we shall focus on the less studied position space. Nevertheless, in Appendix A, besides summarizing the properties of the color noise functions, we discuss the equivalent Fokker-Planck
type \emph{generalized master equations} for the velocity conditional probability $p(v,t\!\mid \!v_{0},0)$ associated
to the \textcolor{black}{GLE}, Eq. (\ref{GLE}).

  In order to derive the corresponding FPE for position, let
  $p(z,t\mid z_{0},v_{0})$ be the \emph{conditional probability distribution} of finding the particle, say at position $z$ at time $t$, given it started to diffuse at $z_{_{0}}$
  with velocity $v_{0}$ at $t=0$. To obtain the FPE for
  the evolution of $p(z,t\mid z_{0},v_{0})$, associated to Eq. (\ref{GLE}), we will use a method originally developed by
  Sancho {\it et al.}\cite{sancho} and applied in many problems\cite{pj,volkov2}.

For a given realization of the noise $\varphi_v(t)$, Eq. (\ref{vGLE}) describes a flow in
$z$--space. The density of this flow evolves in time according to the
 stochastic Liouville equation
\be
\frac{\partial f(z[\varphi_v],t)}{\partial t}=-\frac{\partial}{\partial z}\left[f(z[\varphi_v],t)\,
\frac{dz[\varphi_v]}{dt}\right] ,\label{sle}
\ee
where $f(z[\varphi_v],t)$ is the probability density of the flow.
Taking into account all realizations of $\varphi_v(t)$, the Liouville equation turns into an ordinary SDE\cite{kubo}. As pointed out
by van Kampen \cite{vankampen2,vankampen1}, the probability density of the fixed realization $z$ at
time $t$, $p(z,t\mid z_{0},v_{0})$, can be obtained by averaging the function $f(z[\varphi_v],t)$ over the distribution of the
colored noise, namely
 \be
 p(z,t\mid z_{0},v_{0})=\Big\langle f\big(z[\varphi_v],t\big) \Big\rangle_{_{\!\!\varphi_v}}.
 \label{vanKampen}\ee

 Then, by replacing Eq. (\ref{vGLE}) into Eq. (\ref{sle}), the equation satisfied by $p(z,t\mid z_{0},v_{0})$ reads:
\bea
\frac{\partial\, p(z,t\mid z_{0},v_{0})}{\partial t}&=&-\overline{v}(t)\frac{\partial\, p(z,t\mid z_{0},v_{0})}{\partial z}\nonumber\\
&-&\frac{\partial}{\partial z}\Big\langle f\big(z[\varphi_v],t\big)\,\varphi_v(t)\Big\rangle_{\!\!\varphi_{v}}\!\!.
\label{vk3}
\eea
 Here, the subindex $\varphi_v$ is a remainder that the probability density of the colored noise has to be employed
in the calculation of the average. Since $\varphi_v(t)$ is a zero--mean Gaussian noise, the cross correlation
$\langle f(z[\varphi_v],t)\,\varphi_v(t)\rangle_{_{\!\varphi_v}}$ is
given by the formula of differentiation due to Furutzu\cite{furutzu}, Novikov \cite{novikov} and Donsker\cite{donsker}
\bea
&-&\Big\langle f\big(z[\varphi_v],t\big)\,\varphi_v(t)\Big\rangle_{\!\!\varphi_v}=\nonumber\\
&&\int_{0}^{t}\Big\langle \varphi_v(t)\,\varphi_v(s)\Big\rangle
\left\langle\frac{\partial f(z[\varphi_v],t)}{\partial z[\varphi_v]} \frac{\delta z[\varphi_v]}{\delta\varphi_v(s)}\right\rangle ds ,\nonumber\\
&=&\int_{0}^{t}\Big\langle \varphi_v(t)\,\varphi_v(s)\Big\rangle \!\frac{\partial\, p(z,t\!\mid \!\!z_{0},v_{0})}{\partial z}ds,
\eea
where the functional derivative $\delta z[\varphi_v]/\delta \varphi_v(s)=1$ was obtained from the definitions, Eqs. (\ref{zGLE}) and (\ref{varphiz}), and van Kampen definition, Eq. (\ref{vanKampen}), was used.

Then, carrying out the proper substitutions in Eq. (\ref{vk3}), we finally find that the probability density satisfies the following partial differential equation
\bea
\!\!\!\Big(\frac{\partial\, p(z,t|z_{0},v_{0})}{\partial t}\Big)_{z}&=&-\overline{v}(t)\,\frac{\partial\, p(z,t|z_{0},v_{0})}{\partial z}\nonumber\\
&+&D_{q}(t)\,\frac{\partial^{2}p(z,t|z_{0},v_{0})}{\partial z^{2}},
\label{z-GRFPE}
\eea
where the time dependent diffusion term $D_{q}(t)$ was written as
\be
D_{q}(t)=\int_{0}^{t} \Big\langle \varphi_v(t)\,\varphi_v(s) \Big\rangle\!_{_{R}} \,ds,
\label{functionDq}
\ee
in which $C(t,s)=\big\langle \varphi_v(t)\,\varphi_v(s) \big\rangle\!_{_{R}}$ is the two times correlation of the colored noise response function.
This differential equation is often referred as a Fokker-Planck equation. It is actually a Kramers-Moyal z-space master equation, analogous to the Rayleigh equation in v-space. To emphasize the $v_{0}$ dependence and the fact that the associated SDEs, Eqs. (\ref{vGLE}) and (\ref{zGLE}), constitute the exact solution for the \textcolor{black}{GLE}, we shall refer to Eq. (\ref{z-GRFPE}), together with the z-moments derivatives, Eqs. (\ref{vtilde}) and (\ref{functionDq}), as the \emph{z-space generalized Rayleigh-Fokker-Planck equation} (z-GRFPE). We reserve the acronym GFPE for the master equation associated to the $v_{0}$ averaged $p(z,t\mid z_{0})$ to be obtained in next section.

\section{Time dependent external force: Alternative SDE GFPE for position and velocity.}
\label{sec:II GFPE}

In this section we write equivalent forms of the z-GRFPE, Eq. (\ref{z-GRFPE}), corresponding all to the non-static
GLE, Eq. (\ref{GLE}), and give the analytical solutions for $p(z,t\mid z_{0},v_{0})$ and its initial velocity average
$p(z,t\!\mid\! z_{0})=\langle p(z,t\!\mid\! z_{0},v_{0})\rangle_{v_{0}}$, for any given retardation kernel $\Gamma(t)$
and external force $F(t)$.
\subsection{Solution of Position GFPE}
\label{sec: II.A Position}
 First of all, applying the simple linear transformation
$q(t)=z(t)-\overline{z}(t)$, with the variables
$\overline{z}(t)$ and $\overline{v}(t)=\dot{\overline{z}}(t)$ as defined in previous section,
Eqs. (\ref{vtilde}) and (\ref{ztilde}), Eq. (\ref{z-GRFPE}), reduces to a diffusion--like equation\cite{gnedenko}
\be
\Big(
\frac{\partial \,p(q,t\mid q_{0},v_{0})}{\partial t}\Big)_{q}=
D_{q}(t)\,\frac{\partial^{2}p(q,t\mid q_{0},v_{0})}{\partial q^{2}},
\label{DiffusionGFPE}
\ee
Its solution with an initial condition $\delta(q-q_{_{0}})$ is a Gaussian centered at $q_{_{0}}=0$. In terms of the original
variables, the conditional probability density for a given
realization $z$, starting from $z_{_{0}}$, given an initial velocity $v_{0}$ in a GLE process described by Eq. (\ref{GLE}) is then
\be
p(z,t\mid z_{0},v_{0})\!=\! \frac{1}{\sqrt{2\,\pi\,\sigma_{q}^{^{2}}(t)}}
\exp\left[-\frac{(z-\overline{z}(t))^{2}}{2\,\sigma_{q}^{2}(t)}\right],
\label{chandra}
\ee
 where the standard deviation $\sigma_{q}^{2}(t)=\langle (z-\overline{z}(t))^{2}\rangle\!_{_{R}}$ is defined as
\be
\sigma_{q}^{2}(t)\!=2 \int_{0}^{t}D_{q}(s)\,ds.
\label{sigma2}
\ee

The correlation $C(t,s)$ is evaluated in Appendix A, Eq. (\ref{corfox}). Using it into the expression obtained above for the function
 $D_{q}(t)$, Eq. (\ref{functionDq}), we can readily evaluate
\bea
D_{q}(t)\!\!\!\!&=&\int_{0}^{t}C(t,s)ds,\nonumber\\
&=&\frac{k_{_{\mathrm{B}}}T}{m}\Big[\!\!\int_{0}^{t}\!\!\!\chi_{v}(\mid t-s\mid)ds-\chi_{v}(t)\!\!\int_{0}^{t}\chi_{v}(s)ds\Big]\nonumber\\
&-&\phi_{v}(t)\int_{0}^{t}\phi_{v}(s)ds,
\eea
Using the definitions of the functions $\chi_{z}(t)$ and $\phi_{z}(t)$
\be
D_{q}(t)=\frac{k_{_{\mathrm{B}}}T}{m}\chi_{z}(t)\bigg[1-\chi_{v}(t)\bigg]
-\phi_{v}(t)\phi_{z}(t).
\label{Dq-zGRFP}\ee
A second integration gives the second moment of $q=[z(t)-\overline{z}(t)]$ in terms of the function $\chi(t)$
\be
\sigma^{2}_{q}(t)=\frac{k_{_{\mathrm{B}}}T}{m}\bigg[2\,\chi(t)-
\chi_{z}^{2}(t)\bigg]
-\phi^{2}_{z}(t).
\label{sigmaq-GLE}
\ee
The probability $p(z,t\mid z_{0})$, of finding the particle at position $z(t)$ at time $t$ given that it was initially located at $z_{0}$, independently of the initial velocity, is obtained averaging the conditional
probability $p(z,t\mid z_{0}, v_{0})$ over the $v_{_{0}}$ Maxwellian distribution
 \be
p_{1}(v_{0})\!=\! \frac{1}{\sqrt{2\,\pi\,\frac{k_{_{\mathrm{B}}}T}{m}}}
\exp\left[-\frac{m v_{0}^{2}}{2\,k_{_{\mathrm{B}}}T}\right],
\label{v0-Maxwell}
\ee
A simple integration of the Gaussian in Eq. (\ref{chandra}) gives
\bea
\!\!p(z,t\!\mid\! z_{0})\!&=&\int_{-\infty}^{\infty}p(z,t\mid z_{0}, v_{0})p_{1}(v_{0})dv_{0}\nonumber\\
&=&\!\! \frac{1}{\sqrt{2\,\pi\,\sigma_{z}^{^{2}}(t)}}
\exp\!\!\left[-\frac{\big[z\!-\langle\bar{z}(t)\rangle_{_{\!v_{0}}} \big]^{2}}{2\,\sigma_{z}^{2}(t)}\!\right],
\label{Probfinal}
\eea
where
\bea
\sigma_{z}^{2}(t)&=&\sigma_{q}^{2}(t) +\frac{k_{_{\mathrm{B}}}T}{m}\chi_{_{z}}^{2}(t),\label{sigmaz}\\
&=&2\frac{k_{_{\mathrm{B}}}T}{m} \chi(t) -\phi^{2}_{z}(t).
\eea
Therefore, the diffusion coefficient for $p(z,t\mid z_{0})$ is:
\bea
D_{z}(t)&=&D_{q}(t) + \frac{k_{_{\mathrm{B}}}T}{m}\chi_{z}(t)\chi_{v}(t)\nonumber\\
&=&\frac{k_{_{\mathrm{B}}}T}{m}\chi_{z}(t) -\phi_{v}(t)\phi_{z}(t).
\label{DiffusionzGFPE}\eea
We should point out the fact that, since the particle is free to move in the entire space, $p(z,t\mid z_{0})$ is referred to as the \emph{unbounded} position probability density in one dimension. It is in fact the solution of the exact \emph{generalized Fokker-Planck equation} for position $z$ (z-GFPE)
\bea
\Big(\frac{\partial \,p(z,t\mid z_{0})}{\partial t}\Big)_{z}&=&-\phi_{v}(t)
\frac{\partial\, p(z,t\mid\!z_{0})}{\partial z}\nonumber\\
&+&D_{z}(t)\,\frac{\partial^{2} p(z,t\mid\! z_{0})}{\partial z^{2}},
\label{z-GFPE}
\eea
with the boundary and initial conditions

\bea
p(-\infty,t\mid z_{0})&=&p(\infty,t\mid z_{0})=0,\nonumber\\
p(z,t=0\mid z_{0})&=&\delta(z-z_{0}),
\eea
where
\bea
\langle\bar{v}(t)\rangle_{_{\!v_{0}}}
&=&\langle\dot{\bar{z}}(t)\rangle_{_{\!v_{0}}}=
\phi_{v}(t),\\
\langle\bar{z}(t)\rangle_{_{\!v_{0}}}&=&z_{0}
+\phi_{z}(t),\\
z_{0}&=&\langle\bar{z}(t=0)\rangle_{_{\!v_{0}}}.
\eea

 For an unbounded particle, using these last relationships, it is straightforward to write the
 \textcolor{black}{MSD} with respect to the initial position, denoted as $\sigma^{2}(t)$ without any subscript, i.e.,
 \be
 \sigma^{2}(t)=MSD(t)=\langle\langle [z(t)-z_{0}]^{2}\rangle\!_{_{R}}\rangle_{v_{0}},\ee
we get
 \be
\sigma^{2}(t)=\sigma_{z}^{2}(t) + \phi^{2}_{z}(t)=
2\frac{k_{_{\mathrm{B}}}T}{m} \chi(t).
\label{MSD(t)-F(t)}
\ee
This result can also be obtained directly from Kubo's $v_{0}$-averaged VAC, $c_{v}(t)=(k_{_{\mathrm{B}}}T/m)\chi_{v}(t) $ \cite{kubo,BerneHarp}
\be
\sigma^{2}(t)=2\int_{0}^{t}(t-s)c_{_{v}}(s)ds.
\label{MSD-Kubo}
\ee
The regular definition of the diffusion coefficient commonly related to experimental data or molecular simulations is
\be
D(t)=\frac{1}{2}\frac{d}{dt}\sigma^{2}(t)= \int_{0}^{t}c_{_{v}}(s)ds= \frac{k_{_{\mathrm{B}}}T}{m} \chi_{z}(t).
\label{D(t)-VAC}
\ee
It is interesting that $D(t)$ so defined is not explicitly dependent of the external force, while the coefficient directly associated to $p(z,t\mid z_{0})$, namely
$D_{z}(t)=D(t) - \phi_{v}(t) \phi_{z}(t)$, does depend on $F(t)$.
It is also common to find the diffusion coefficient defined as an extension of the SE result $\sigma^{2}(t)=2\widetilde{D}t$, i.e.
\be
\widetilde{D}(t)=\frac{\sigma^{2}(t)}{2 t}= \frac{1}{t}\int_{0}^{t}D(s)ds= \frac{k_{_{\mathrm{B}}}T}{m} \frac{\chi(t)}{t}.
\label{Dtilde(t)}
\ee
$\widetilde{D}(t)$ is then the time average of $D(s)$ in the interval $(0,t)$. The inconvenience of the use of $\widetilde{D}(t)$ is discussed in next section, Fig. (\ref{Fig1-Dfree}).

For completeness, note that, with a simple change of variables the z-GFPE, Eq. (\ref{z-GFPE}), can be written in the usual form of a diffusion equation. Defining the relative position $Z(t)=z(t)-\langle\bar{z}(t)\rangle_{_{\!v_{0}}}
=z(t)-z_{0}-\phi_{z}(t)$
and using the relationship
\bea
\Big(\frac{\partial\, p(Z,t\!\!\mid \!\!Z_{0})}{\partial t}\Big)_{Z(t)}&=&\Big(\frac{\partial\, p(z,t\!\!\mid \!\!z_{0})}{\partial t}\Big)_{z(t)}\nonumber\\
&+&\langle\dot{\bar{z}}(t)\rangle_{_{\!v_{0}}}\Big(\frac{\partial p(z,t\!\!\mid \!\!z_{0})}{\partial z}\Big)_{t},
\label{dPdt-x}
\eea
 we get the velocity--independent generalized diffusion equation (GDE)
\be
\Big(\frac{\partial\, p(Z,t\!\!\mid \!\!Z_{0})}{\partial t}\Big)_{Z(t)}
=D_{z}(t)\,\frac{\partial^{2} p(Z,t\mid\!\!Z_{0})}{\partial Z^{2}},
\label{GLE-DiffusionEq}
\ee
with a time dependent diffusion term $D_{z}(t)$ given by Eq. (\ref{DiffusionzGFPE}). According to this, it is identical to $D(t)$ only for zero external force. In that case, the common Sutherland-Einstein limit $D_{z}(t)=D(t) \rightarrow D_{0}$ is obtained in the Smoluchowski limit where $\alpha \chi_{z}(t)\rightarrow 1 $.

This generalized unbounded diffusion equation satisfies the initial and boundaries conditions: $Z_{0}=0$ and
\bea
&& p(Z,t=0\mid Z_{0}=0)=\delta(Z),\nonumber\\
&& p(-\infty,t\mid Z_{0})= p(\infty,t\mid Z_{0})=0.
\eea
\subsection{Solution of velocity GFPE}
\label{sec:II.B Velocity}
The z-space master equation, Eq. (\ref{z-GRFPE}), gives the probability density $p(z,t\!\!\mid\!\! z_{0},v_{0})$. It is associated to the process defined by the GLE, with $v(t)=\dot{z}(t)$ given by Eq. (\ref{vGLE}), which can be written as
 \be
 v(t)= v_{0}\chi_{v}(t) + \varphi(t),
\label{v-varphi}\ee
with the total internal plus external force response function defined as
\be
\varphi(t)=\varphi_{v}(t)+\phi_{v}(t),
\ee
in which, the random velocity term is given by a colored noise function $\varphi_v(t)$ satisfying $\langle \varphi_{v}(t)\rangle_{\!R}=0$ and the external force was assumed to be time dependent but, constant in space. Therefore the moments of the {\it v}-space probability density $p(v,t\mid v_{0})$ should be closely related to those of $p(z,t\mid z_{0},v_{0})$. In fact, using Chandrasekhar's
argument\cite{chandra,mcquarrie}, we show in Appendix A that $p(v,t\mid v_{0})$ can be written as a normal Gaussian distribution, see Eq. (\ref{chandravelocity}).
Consequently, it satisfies a generalized master equation identical to the velocity Rayleigh equation for the LE,
\bea
\!\!\!\! \Big(\frac{\partial p(v,t\mid v_{0})}{\partial t}\Big)_{\!\!v}&=&-\dot{\overline{v}}(t)\,\frac{\partial \,p(v,t\mid v_{0})}{\partial v}\nonumber\\
&+&D_{u}(t)\,\frac{\partial^{2}p(v,t\mid v_{0})}{\partial v^{2}},\label{v-GRFPE}
\eea
but with the first and second moments generalized as
\be\overline{v}(t)= v_{0}\chi_{v}(t) + \phi_{v}(t),
\label{vbarFt}
\ee
\be
\sigma_{u}^{2}(t)=\Big\langle \varphi_v^{2}(t)\Big\rangle= \frac{k_{_{\mathrm{B}}}T}{m}
\bigg[1-\chi^{2}_{v}(t)\bigg] -\phi_{v}^{2}(t).
\label{sigmau2Ft}
\ee
We shall refer to Eq. (\ref{v-GRFPE}), together with (\ref{vbarFt}) and (\ref{sigmau2Ft}) as the generalized velocity Rayleigh-Fokker-Planck equation (v-GRFPE).

With the definition of the total response function $\varphi$, Eq. (\ref{v-varphi}), it is straightforward to obtain the relationships
\bea
 \dot{v}(t)&=&-\beta(t)v(t) +\beta(t)\varphi(t)+\dot{\varphi}(t),\\
\dot{\overline{v}}(t)&=&-\beta(t)\overline{v}(t) + \beta(t)\phi_{v}(t) + \dot{\phi}_{_{v}}(t),
\eea
where
\bea
\beta(t)
&=&-\frac{\dot{\overline{v}}(t)-\dot{\phi}_{_{v}}(t)}
{\overline{v}(t)-\phi_{v}(t)},\nonumber\\
&=&-\dot{\chi_{v}}(t)/\chi_{v}(t).
\eea
Substituting this in the v-GRFPE, Eq. (\ref{v-GRFPE}), we then get the standard form of the\emph{ velocity generalized Fokker-Planck } equation (v-GFPE)\cite{adelman1,fox1}
\bea
\!\!\!\frac{\partial\, p(v,t\mid v_{0})}{\partial t}&=&
\beta(t)\frac{\partial}{\partial v}\big[v_{\mathrm{drift}}(t)
p(v,t\mid v_{0})\big]\nonumber\\
\!\!\!&+&\,\,D_{v}^{\mathrm{A}}(t)\,\frac{\partial^{2}p(v,t\mid v_{0})}{\partial v^{2}}.\hspace{30pt}
\label{v-GFPE-F(t)}
\eea
In which the drift and diffusion terms are
\be
v_{\mathrm{drift}}(t)=v(t) + (\Delta v)_{_{F}},
\ee
and
\be
D_{v}^{\mathrm{A}}(t) = D_{u}(t) + \beta(t)\sigma_{u}^{2}(t)=\frac{k_{_{\mathrm{B}}}T}{m}\beta(t)+(\Delta D)_{_{F}},
\ee
where the shift in the drift and dispersion coefficients due to the presence of the time dependent external force can be written in Adelman's notation \cite{adelman1} as:
\bea
(\Delta v)_{_{F}}&=&-\chi_{v}(t)
\frac{d}{dt}\Big[\frac{\phi_{v}(t)}{\chi_{v}(t)}\Big],\\
(\Delta D)_{_{F}}&=&
-\frac{1}{2}\chi^{2}_{v}(t)
\frac{d}{dt}\Big[\frac{\phi_{v}(t)}{\chi_{v}(t)}\Big]^{2}.
\eea
In the case of a free diffusing particle, $\phi_{v}(t)=0$, and this reduces to the standard form of the Fokker-Planck equation, Eq. (\ref{ME-Adelman}), as derived by Adelman\cite{adelman1}. A similar result was suggested by H\"{a}nggi and Talkner\cite{hanggiTalkner} but they omitted the last diffusive term, since their FDT did not contain the extra force term pointed out in Appendix A, Eq. (\ref{App-FDTheorem}).

The velocity probability density $p_{1}[v(t)]=p(v,t)$, irrespective of the initial velocity, is obtained averaging over $p_{1}[v_{0}]$, the distribution of the $v_{0}$, i.e
\be
p_{1}[v(t)]= \,\!\!\!\!\int_{-\infty}^{\infty}\!\!\!p(v,t\mid v_{0})p_{1}[v_{0}]dv_{0}.
\ee
Using the Gaussian distributions with the proper standard deviations, Eq. (\ref{chandravelocity}) for $p(v,t\mid v_{0})$, and Eq.(\ref{v0-Maxwell}) for $p_{1}[v_{0}]$,
we obtain
\be
p_{1}[v(t)]=\! \frac{1}{\sqrt{2\,\pi\,\sigma_{v}^{2}}}
\exp\left[-\frac{[ v(t)-\phi_{v}(t)]^{2}}{2\,\sigma_{v}^{2}}\right],
\label{App-p(v,t)}
\ee
\bea
\sigma_{v}^{2}&=&\sigma_{u}^{2}
+\frac{k_{_{\mathrm{B}}}T}{m}\chi_{v}^{2}(t),\nonumber\\
&=&\frac{k_{_{\mathrm{B}}}T}{m}-\phi_{v}^{2}(t).
\label{App-sigmav}
\eea
Since $\phi_{v}(t)=0$ at $t=0$, $\sigma_{v}^{2}(0)$ reduces to
$k_{_{\mathrm{B}}}T/m$ , as expected. However, for $t>0$ the $p_{1}[v(t)]$ has a drifting term due to the effect of the external force. Nevertheless, the velocity distribution is stationary, since the average of $v^{2}(t)$ over the probability $p_{1}[v(t)]$ is
\be
\langle v^{2}(t)\rangle_{v}=\frac{k_{_{\mathrm{B}}}T}{m},\ee
 This is consistent with the physical equipartition initial condition $\langle v_{0}^{2}\rangle_{_{\!v_{0}}}
=k_{_{\mathrm{B}}}T/m$, described in Appendix A, Eq.(\ref{App-conditions}).
\subsection{Alternative view of the SDE}
\label{sec:II.C Alternative}
Equation (\ref{z-GRFPE}) is formally associated to the SDE
\be
 dz(t)=\overline{v}(t)\,dt + \varphi_{v}(t)\,dt.
 \label{originalSDE-GLE}
 \ee
However, since Eq. (\ref{z-GRFPE}) is a well defined master equation, by inspection we can reinterpret it as resulting from the direct
application of Ito's lemma\cite{gardiner} to the following SDE:
 \be
 dz(t)=\overline{v}(t)\,dt + \sqrt{2D_{q}(t)}\,\xi(t)\,dt.
 \label{newSDE-PJ-GLE}
 \ee
It has the same form of the drift term, but now the random contribution is given in terms of the standard Gaussian $\delta$--correlated white noise $\xi(t)$. It is weighed
by a time dependent diffusion coefficient $D_{q}(t)$, defined by Eq. (\ref{Dq-zGRFP}). This result justifies the use of H\"{a}nggi's type SDE, Eq. (\ref{eqzeta}) discussed in Appendix A, with $\zeta(t)=z(t)$, $a(t)=\overline{v}(t)=v_{0}\chi_{v}(t) + \phi_{v}(t)$, $b(t)=D_{q}(t)$ and, Eq. (\ref{FPE-Rayleigh}) with $p(\zeta,t)=p(z,t\mid z_{0},v_{0})$.
 This argument is also true for Eq. (\ref{z-GFPE}), which results from
\be
 dz(t)=\phi_{v}(t)dt
  + \sqrt{2D_{z}(t)}\,\xi(t)\,dt.
\label{newSDE-WOR-GLE}
\ee
This result corresponds to the use of Eq. (\ref{eqzeta}), with $\zeta(t)=z(t)$, $a(t)=\phi_{v}(t)$, $b(t)=D_{z}(t)$, and Eq. (\ref{FPE-Rayleigh}) with $p(\zeta,t)=p(z,t\mid z_{0})$.

Even though the dynamics of the SDEs, Eqs. (\ref{originalSDE-GLE}), (\ref{newSDE-PJ-GLE}) and (\ref{newSDE-WOR-GLE}), are different, they are equivalent
in the sense that their statistical properties should be identical. In fact, from Eq. (\ref{newSDE-PJ-GLE})
\bea
z(t)-z_{_{0}} &=& \int_{0}^{t}\overline{v}(s)\, ds +\int_{0}^{t}\!\!\sqrt{2D_{q}(s)}\,\xi(s)\,ds,
\eea
 and recalling that $\langle\xi(t)\rangle=0$ and $\langle\xi(t)\xi(s)\rangle=\delta(t-s)$, we have, after taking the $\xi(s)$ distribution average
 \bea
\Big\langle z(t)\Big\rangle_{\xi}& =& \!\!z_{_{0}}+\int_{0}^{t}\overline{v}(s)\, ds=\Big\langle z(t)\Big\rangle_{R},\label{newztilde}\\
\Big\langle z(t)^{2}-\overline{z}(t)^{2} \Big\rangle_{\xi}
&=&2\int_{0}^{t}D_{q}(s)\,ds=\sigma_{q}^{2}(t),\label{newsigma}
\eea
in agreement with Eqs. (\ref{chandra}) and (\ref{sigma2}).
Therefore, equation (\ref{newSDE-PJ-GLE}) also describes the position of a 1--D Brownian particle evolving with time in the GLE picture. This new SDE is simpler than the original, having the form of the Smoluchowski equation, but is valid for all values of the static friction coefficient and not limited to the condition imposed by the high friction limit approximation. It reduces to the static LE for large $\lambda$. In the same manner, Eq. (\ref{newSDE-WOR-GLE}) is in agreement with Eqs. (\ref{Probfinal}) and (\ref{sigmaz}) with $\langle\bar{z}(t)\rangle_{_{\!v_{0}}}-z_{0}=\phi_{z}(t)$.
Hence, Eqs. (\ref{newSDE-PJ-GLE}) and (\ref{newSDE-WOR-GLE}) are alternative Langevin processes that describe the position $z(t)$ of a non-markovian particle with probability densities $p(z,t\!\mid \!z_{0}, v_{0})$ and $p(z,t\mid z_{0})$ respectively, related to the colored noise process, Eq.(\ref{originalSDE-GLE}). Their simplicity could be useful in the description of more complex systems. We point out that the velocity distributions associated to the above three SDE must be different in order to be consistent with the scaled $z$ space.
\section{Analytical solution of the position z-GRFPE and z-GFPE for a fluid under a constant force.}
\label{sec:III Analytical}
To describe a real viscoelastic fluid one needs to specify its bulk long time self diffusion constant $D_{0}$, besides other molecular parameters as the mass $m$ and the temperature $T$. The bulk static friction coefficient is then given by the Sutherland-Einstein relationship $ \gamma= k_{_{\mathrm{B}}}T/D_{0}$, which is equivalent to the definition $D_{_{0}}\alpha=k_{_{\mathrm{B}}}T/m$ with $\alpha=\gamma/m$.
To solve the GLE, the friction kernel $\Gamma(t)$ needs to be specified. As a first order approximation, we consider the standard exponential decaying kernel \be
\Gamma(t)=\alpha\,\lambda\,\mbox{e}^{-\lambda \,t}.\label{Gamma}
\ee

This is convenient because it depends only on two parameters, and renders simple analytical relationships. Here, the frequency $\alpha$ measures the magnitude of the static friction, while $1/\lambda$ is the memory
relaxation time\cite{gardiner,liu}.
For this approximation, the characteristic Green function $\chi_{v}(t)$ and the auxiliary functions $\chi_{z}(t)$, and $\chi(t)$, Eqs. (\ref{chihat}), (\ref{chiz1}, \ref{chiF1}) respectively, can be evaluated analytically by solving the inverse Laplace
transform of $\widehat{\Gamma}(k)$\cite{dufty},
\bea
\chi_{v}(t)\!\!
&=&\!\!\mbox{e}^{-\lambda\,t/2} \left[\cos\big(\omega\,t\big)+\frac{\lambda}{2\,\omega}\, \sin\big(\omega\,t\big)\right],\label{chiv}\\
\alpha \chi_{z}(t)\!\!
&=&\!\!1\!\!-\mbox{e}^{-\lambda t/2}\!\! \left[\cos\big(\omega\,t\big) \! + \!\frac{\lambda\!-\!2\,\alpha}{2\,\omega}
\sin\big(\omega\,t\big)\right]\!\!,\label{chiz}\\
\alpha\chi(t)\!\!&=&\!\!\!t+ \!\!\frac{\lambda-\alpha}{\alpha\,\lambda}\left[ \mbox{e}^{-\lambda t/2}\!\cos\big(\omega\,t\big)-1\right]\nonumber\\
&+&\frac{\lambda\!-\!3\,\alpha}{2\,\alpha\,\omega}\mbox{e}^{-\lambda t/2}\sin\big(\omega\,t),
\label{chiF}
\eea
where $\omega$ is an auxiliary frequency, $2\,\omega=\sqrt{\lambda\,(4\,\alpha-\lambda)}$.

For the case of a constant external force $\phi_{z}(t)=a_{_{\mathrm{F}}}\,\chi(t)$ and $\phi_{v}(t)=a_{_{\mathrm{F}}}\,\chi_{_{z}}(t)$, where $a_{_{\mathrm{F}}}=v_{_{\mathrm{F}}}\alpha=F/m$ measures the magnitude of the force. The parameters for the probability density $p(z,t\mid z_{0},v_{0})$ in the z-GRFPE are
\bea
\overline{z}(t)&=&z_{0}+v_{_{\!0}} \,\chi_{z}(t)
+a_{_{\mathrm{F}}}\chi(t),
\label{ztilde-Fconst}\\
\sigma^{2}_{q}(t)&=&D_{0}\alpha\big[2\,\chi(t)-
\chi_{z}^{2}(t)\big]
-a_{_{\mathrm{F}}}^{2}\,\chi^{2}(t),
\label{sigmaq-zGRE}\\
D_{q}(t)&=&D_{0}\alpha\chi_{z}(t)\big[1-\chi_{v}(t)\big]
- a_{_{\mathrm{F}}}^{2}\chi_{z}(t)\chi(t).
\label{D-zGRE-Fconst}
\eea
While the parameters for $p(z,t\mid z_{0})$ in the z-GFPE are obtained as
\bea
\langle\bar{z}(t)\rangle_{_{\!v_{0}}}\!&=&\!z_{0}
+a_{_{\mathrm{F}}}\chi(t),\label{zave-Fconst}\\
\sigma_{z}^{2}(t)&=&2D_{0}\alpha \chi(t) -a_{_{\mathrm{F}}}^{2}\chi^{2}(t),\label{Sigmaz-Fconst}\\
D_{z}(t)&=&D_{0}\alpha\chi_{z}(t) -a_{_{\mathrm{F}}}^{2}\chi_{z}(t)\chi(t).\label{Dz-Fconst}
\eea
 A central result is that the MSD parameters are independent of the applied external force
\bea
\!\!\sigma^{2}(t)
&=&2 D_{0} \alpha \chi(t),\label{sigmaMSD-Fconst}\\
\widetilde{D}(t)&=&\!\!D_{0} \frac{\alpha \chi(t)}{t},\\
\!\!D(t)\!\!&=&\!\!D_{0}\!-D_{0}\mbox{e}^{-\lambda
 t/2}[\cos\omega\, t \! +
 \!\frac{\lambda\!-\!2\,\alpha}{2\,\omega}\sin\omega\,
 t].\label{D(t)-Fconst}
\label{D(t)tilde-Fconst}
\eea
The moments for the velocity v-GRFPE probability density $p(v,t\mid v_{0})$ are
\bea
\overline{v}(t)&=& v_{0}\chi_{v}(t) + a_{_{\mathrm{F}}}\chi_{z}(t),
\label{vbar-Fconst}\\
\sigma_{u}^{2}(t)&=&\frac{k_{_{\mathrm{B}}}T}{m}
\big[1-\chi^{2}_{v}(t)\big]-a_{_{\mathrm{F}}}^{2}\chi_{z}^{2}(t).\label{sigmau2-Fconst}\\
D_{u}(t)&=& -D_{0}\alpha\dot {\chi}_{v}(t)\chi_{v}(t) -a_{_{\mathrm{F}}}^{2}(t)\,\chi_{v}(t)\chi_{z}(t),
\label{Du-Fconst}
\eea
while those for $p_{1}[v,t]$ they should be
\bea
\langle\bar{v}(t)\rangle_{_{\!v_{0}}}
&=& a_{_{_\mathrm{F}}}\chi_{z}(t),\label{v-ave-Fconst}\\
\sigma_{v}^{2}&=&\frac{k_{_{\mathrm{B}}}T}{m}-a_{_{\mathrm{F}}}^{2} \chi_{z}^{2}(t),\label{sigmav-Fconst}
\eea
\begin{figure}[ht]
\includegraphics[width=0.48\textwidth]{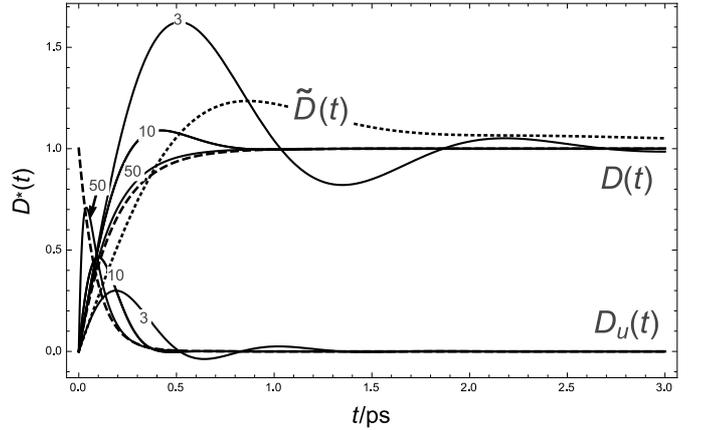}
\caption{The free diffusion reduced position and velocity fluctuation coefficients $D^{*}(t)=D(t)/D_{0}$ and $D^{*}_{u}(t)=D_{u}(t)/(D_{0}\,\alpha^{2})$
versus time $t$ for an argon--like fluid. Solid lines are the GLE prediction for several $\lambda$ values. Dashed curves correspond to LE and superimpose to the GLE curve for $\lambda>50\,\mathrm{ps}^{-1}$. Also shown is the apparent $\widetilde{D}(t)$ function for $\lambda=3\,\mathrm{ps}^{-1}$ (dotted line).}
\label{Fig1-Dfree}
\end{figure}

In Fig. (\ref{Fig1-Dfree}) we compare the behavior of the fluctuation coefficients for position, $D(t)$, Eq. (\ref{D(t)-Fconst}), and velocity, $D_{u}(t)$, Eq. (\ref{Du-Fconst}), as a function of time. The figure is for $a_{_{\mathrm{F}}}=0$, but finite values of the force only affects the values of the velocity $D_{u}(t)$. The GLE results
are shown for several values of $\lambda$.
Particularly, for a very large $\lambda>50$, the curves match the classical LE (dashed curves), given by Eqs. (\ref{app-DvLE}) and (\ref{app-DzLE}) of Appendix B. For smaller $\lambda$, oscillations are present and important discrepancies with LE are observed
at times of the order of $1/\alpha$. As $t$ approaches zero, the ballistic behavior is obtained. We can notice that the time scale of the velocity is much shorter than that of the position. It is an indication that the
velocity field attains an earlier canonical distribution.
 Also shown is the apparent $\widetilde{D}(t)$ function for the low $\lambda=3\,\mathrm{ps}^{-1}$ (dotted line). We can see that $\widetilde{D}(t)$ depicts only the time average of the diffusivity $D(t)$ for the same $\lambda$. At low times it misses the important oscillations present in the actual $D(t)$. It requires extremely large times to converge to the correct limit $D_{0}$. Therefore its use is misleading.
\section{ GLE Mean First Passage and Persistence Time.}
\label{sec:IV Dynamical}
So far we have dealt with unbounded particles. In this section we shall take a look at the diffusion of constrained particles. First passage time distributions have been extensively studied due to its related applications to diffusion controlled kinetics and biological processes \cite{szabo,gitterman,szabo2,hu-chen-berne}.
Starting from the high friction limit diffusion equation, many physically valid boundary conditions have been applied, leading to well known analytical results. Based on the simple representation obtained in previous section, here we shall develop the GLE theory of first passage times of constrained viscoelastic fluid particles, under an external force. So, the z-GFPE, Eq. (\ref{z-GFPE}), must be solved with the appropriate boundary conditions for one and two absorbing bounding barriers.
\subsection{Diffusion Next to a Single Absorbing Barrier}
\label{sec:FPT}
 Let us first consider a fluid particle initially at $z_{0}$, diffusing under a force $F(t)$, but constrained in the $z$ direction by an absorbing barrier located at position $b>z_{0}$. The absorbing property of the barrier requires that the boundary condition $p(b,t\mid z_{0})=0$ is satisfied. This can be fulfilled representing the bounded probability by the standard combination of unbounded $p^{0}(z,t\mid z_{0})$ functions \cite{mcquarrie}, namely
\be
p(z,t\mid z_{0})= p^{0}(z,t\mid z_{0})- p^{0}(2b-z,t\mid z_{0}).
\ee
Integration over $z$ gives, after some algebra, the survival probability
 \bea
G(z_{0},t)&=&\int_{-\infty}^{b}p(z,t\mid z_{0})\,dz,\nonumber\\
&=& \mbox{erf}\Big[\frac{Z_{b}(t)}{\sqrt{2 \sigma^{2}_{z}(t)}}\Big],
\label{Gz0t-1wall}
\eea
where $Z_{b}(t)\!=\!b -z_{0}-\phi_{z}(t)$, and $\mbox{erf}[z]$ is the error function of $z$.

Defining $h(t,z_{0})\!\!=\!-dG(t,z_{0})/dt$ as the \emph{first passage time distribution}, we get
\bea
h(t,z_{0})&=& \frac{2\big[Z_{b}(t)D_{z}(t)+\phi_{v}(t)\sigma^{2}_{z}(t) \big]}{\sqrt{2\pi }\sigma^{3}_{z}(t)}
\nonumber\\
&\times& \exp{\Big[-\frac{Z_{b}^{2}(t)}{2\sigma^{2}_{z}(t)}\Big]}.
\label{h-F(t)}
\eea
In the absence of an external force, this reduces to
\bea
[h(t,z_{0})]_{\mathrm{F}=0}&=& \frac{\alpha\chi_{z}(t)(b-z_{0})}{\sqrt{4\pi D_{0}} \big[\alpha\chi_{_{\mathrm{F}}}(t)\big]^{3/2}}
\nonumber\\
&\times&\exp{\Big[-\frac{(b-z_{0})^{2}}{4D_{0} \alpha \chi_{_{\mathrm{F}}}(t)}\Big]}.
\label{h-F0}
\eea

Equation (\ref{h-F(t)}) is an exact closed relationship for the dynamic generalized Langevin equation of a viscous fluid under a time-dependent space-fixed finite external force. For a time-constant force and an exponential kernel, it is analytical, since the functions $\chi_{v}(t)$, $\chi_{_{\mathrm{F}}}(t)$, $\phi_{v}(t)$, $\sigma^{2}_{z}(t)$, and $D_{z}(t)$ were given analytically in previous sections.
Particularly, in the high friction or Smoluchowski limit ($\lambda\rightarrow\infty$, $\alpha\rightarrow\infty$, and $D_{z}(t)\rightarrow D_{0}$), Eq. (\ref{h-F(t)}) becomes for a constant force
\bea
[h(t,z_{0})]_{_{\mathrm{SE}}}&=& \frac{b-z_{0} + v_{_{\mathrm{F}}} t}{\sqrt{4\pi D_{0}}\, t^{3/2}}\nonumber\\
&\times& \exp{\Big[\!\!-\frac{\!\!(b-z_{0}-v_{_{\mathrm{F}}}t)^{2}}{4 D_{0} t}\Big]},
\label{h-Fconst}
\eea
where, as we defined before, $v_{_{\mathrm{F}}} t = F D_{0} t/k_{_{\mathrm{B}}}T $. Note that Eq. (\ref{h-Fconst}) differs slightly from the Smoluchowski result for a constant force obtained by Hu {\it et al.}\cite{hu-chen-berne} where the drift contribution is missing in the numerator. Additionally, our $\sigma^{2}_{z}$ contribution contains a new term, quadratic in the force.
\subsection{Diffusion Between Two Absorbing Barriers}
\label{sec:Slab}
Another physical situation widely studied in the SE limit corresponds to the self -diffusion of a viscous fluid within an absorbing virtual slab of molecular dimensions.
Let $z=0$ and $z=L$ be the position of two absorbing boundaries (A--A) of a virtual slab of length $L$. The absorbing boundary conditions require that $p(0,t\mid z_{0})=p(L,t\mid z_{0})=0$. For a time dependent drift velocity $\phi_{v}(t)$, due to the external force $F(t)$, the solution of Eq. (\ref{z-GFPE}), can be obtained by separation of variables \cite{mcquarrie,gitterman}. Using the initial condition $p(z,t=0\mid z_{0})=\delta(z-z_{0})$ , we obtain
\bea
\!\!\!p(z,t\mid z_{0})
\!&=&\!\frac{2}{L}\sum_{n=1}^{\infty}
\!\sin(\frac{n\,\pi}{L}\!z)\sin(\frac{n\,\pi}{L}\!z_{0})\!\nonumber\\
&\times&\!\exp\Big[\frac{\mathrm{Pe}(t)z-\mathrm{Pe}^{0}z_{0}}{L}\!-\!\frac{D_{0}}{L^{2}}
T_{n}(t)\,\Big],
\label{pzz0t-F(t)}
\eea
where we introduced the dimensionless time dependent P\'eclet number,
 \bea
\mathrm{Pe(t)}&=&\frac{\phi_{v}(t)L}{2\,D_{z}(t)},\\
&=&\frac{\mathrm{Pe}^{0}}{1-\big(\frac{2\mathrm{Pe}^{0}}{L}\big)^{2}D_{0}\alpha \chi(t)}.
\eea
which measures the importance of the mass transfer due to the external drifting force, relative to that due to diffusion.
Here $\mathrm{Pe^{0}}=\mathrm{Pe} (t=0)=v{_{F}}L/2\,D_{0}$. The function $T_{n}(t)$ is a re-scaled time given by
\bea
\!\!\!\!T_{n}(t)\!\!\!&=&\!\!\frac{[n^{2}\,\pi^{2}+\mathrm{Pe}^{2}(t)]}
{\mathrm{Pe}^{2}(t)}
\!\int_{0}^{t}
\mathrm{Pe}^{2}(s)\frac{D_{z}(s)}{D_{0}}\,ds,\\
\!\!\!\!&=&\!\!\!-
\frac{L^{2}[n^{2}\,\pi^{2}\!\!+\!\mathrm{Pe}^{2}(t)]}
{4D_{0}\mathrm{Pe}^{2}(t)}
\ln\!\Big[1\!\!-\!\frac{2(\mathrm{Pe}^{0})^{2}\sigma^{2}{(t)}}{L^{2}}\Big]
\label{tn(t)}.
\eea

 The \emph{survival probability} for a
particle to remain in this region
having started to diffuse at $z_{0}>0$ is given by
\be
G(z_{0},t)=\int_{0}^{L}p(z,t\mid z_{0})\,dz.
\label{Gz0tdefinition}
\ee
A direct integration gives
\bea
G(z_{0},t)
&=&2\!\sum_{n=1}^{\infty}
\frac{n\,\pi\left[1-\mathrm{e}^{\mathrm{Pe(t)}}
(-1)^{n}\right]}
{\,n^{2}\,\pi^{2}+\mathrm{Pe^{2}(t)}}\!\sin(\frac{n\,\pi}{L}
\!z_{0})\nonumber\\
&\times&\exp\Big[-\frac{\mathrm{Pe^{0}}}{L}z_{0}-
\frac{D_{0}}{L^{2}}T_{n}(t)\,\Big].
\label{Gz0t}
\eea
Note that using the definition of $G(z_{0},t)$ in the backward version of Eq. (\ref{z-GFPE}), we find that the survival probability $G(z_{0},t)$ must obey the differential equation
\bea
\!\!\frac{\partial G(z_{0},t)}{\partial t} \!&=&\!\phi_{v}(t)\,
\frac{\partial G(z_{0},t)}{\partial z_{0}}\!+\!D_{z}(t)\,\frac{\partial^{2} G(z_{0},t)}{\partial z_{0}^{2}},\nonumber\\
G(0,t)&=&G(\,L,t)=0,\nonumber\\
G(z_{0},0)&=&1,
\label{S-GFPE}
\eea
whose solution is just Eq. (\ref{Gz0t}).

The \emph{distribution of the first passage times}, $h(z_{0},t)$ can be evaluated using the fact that it is equal to the net flux reaching the absorbing boundaries
\bea
\!\!h(z_{0},t)
\!\!\!&=&\!\!-\frac{d G(z_{0},t)}{dt},\nonumber\\
\!\!\!&=&\!\!D_{z}(t)\!\Big[\Big(\!\frac{\partial p(z,t\!\mid\!\!z_{0})}{\partial z}\Big)_{\!\!L}\!\!\!-\!\Big(\frac{\partial p(z,t\!\mid\!\!z_{0})}{\partial z}\Big)_{\!\!0}\Big],
\eea
then
\bea
h(z_{0},t)\! &=&2\!\sum_{n=1}^{\infty} \frac{n\,\pi D_{z}(t)}{L^{2}}
 \left[1\!-\!\mathrm{e}^{\mathrm{Pe(t)}}(-1)^{n}\right]\nonumber\\
 &\times&\sin\!\Big(\!\frac{n\,\pi}{L}\!z_{0}\!\Big)\exp\!\Big[\!\!-\!\!\frac{\mathrm{Pe^{0}}} {L}z_{0}\!-\!\!\frac{D_{0}}{L^{2}}
  T_{n}(t)\!\Big]\!.
 \label{hzt-GLE}
 \eea
The \emph{persistence probability} $P(t)$ that the particle is still in the slab $(0,L)$ at time t, irrespective of its initial position, is obtained averaging the survival probability over the distribution of initial positions
 \be
 P(t)=\langle G(z_{0},t)\rangle_{z_{0}}=
 \int_{_{0}}^{^{L}}G(z_{0},t)\,g(z_{0})\,dz_{0},
 \label{PtGLE}
 \ee
where $g(z_{0})$ is the one-particle radial distribution function of the fluid, normalized in the slab $[0,L]$. In the virtual layer model simulations \cite{liu,CLOR} the external mean force is assumed to be a constant within the slab, so the potential of mean force is $W(z_{0})\cong-Fz_{0}$ and in terms of the parameter $\mathrm{Pe}^{0}$
\be
g(z_{0})=\frac{1}{L}N_{g}\exp\Big[\frac{2\mathrm{Pe}^{0}}{L}z_{0}\Big].
\ee
With $N_{g}=2\mathrm{Pe}^{0}/\big[\exp{(2\mathrm{Pe}^{0}})-1\big]$ as the normalization factor.
Integrating over $z_{0}$, the persistence probability in the absorbing slab is
\bea
\!\!\!\!\!P(t)\!\!&=&\!\!2 N_{g}\sum_{n=1}^{\infty}
\!n^{2}\,\pi^{2}\!
\frac{1\!-\!(-1)^{n}\mbox{e}^{\mathrm{Pe^{0}}}}
{\,n^{2}\,\pi^{2}+\mathrm{Pe^{0}}^{2}}\nonumber\\
&\times&\frac{1\!-\!(-1)^{n}\mbox{e}^{\mathrm{Pe}(t)}}
{\,n^{2}\,\pi^{2}+\mathrm{Pe}^{2}(t)}
\exp\Big[-\frac{D_{_{0}}}{L^{2}}T_{n}(t)\Big].
\label{P(t)-F(t)}
\eea
The \emph{mean first passage time} $t_{_{\mathrm{MFP}}}(z_{0})$ (MFPT) is the first moment of $h(z_{0},t)$\cite{gardiner,ColmenaresPRE,worpj}. \be
t_{_{\mathrm{MFP}}}(z_{0})=\int_{0}^{\infty}t\,h(z_{0},t)dt
=\int_{0}^{\infty}G(z_{0},t)dt.
\label{MFPT-GLE-F(t)}\ee
Its average over the distribution of the initial positions is called the \emph{persistence time}\cite{worpj}, denoted by $\tau$. For absorbing boundaries it is also referred to as the \emph{mean exit time}. In terms of $P(t)$, $\tau$ is given by\cite{ColmenaresPRE,worpj}:
\be
\tau=\langle t_{_{\mathrm{MFP}}}(z_{0}) \rangle_{z_{0}}= \int_{0}^{\infty}P(t)\, dt.
\label{tau-GLE-F(t)}
\ee
That is, the persistence time is just the normalization constant of the \emph{persistence probability}.

The mean square displacement for particles persisting in the absorbing slab at time $t$, $\sigma^{2}_{L}(t)$, is a bounded function. Formally, it should be obtained integrating $(z(t)-z_{0})^{2}$ over $z(t)$ and $z_{0}$ with a bounded probability distribution given by Eq. (\ref{pzz0t-F(t)}). However, another simpler expression can be computed in terms of the unbounded diffusion coefficient and the survival probability $P(t)$
\be
\sigma^{2}_{L}(t)=2\int_{0}^{t}D(s)P(s)ds.
\label{MSD-L}
\ee
\subsection{A-A Diffusion for Small Pe Numbers }
\label{sec:Slab-lowPe}
P\'eclet numbers, so defined above, are in general nonlinear functions of time, which depend on the functional form of the external force $F(t)$. Therefore, the time integrations needed to evaluate $t_{_{\mathrm{MFP}}}(z_{0})$, its average $\tau$ and $\sigma^{2}_{L}(t)$, have to be carried out numerically according to the exact GLE closed relationships given by Eqs. (\ref{MFPT-GLE-F(t)}), (\ref{tau-GLE-F(t)}), and (\ref{MSD-L}), for a given value of $\lambda$.

To obtain analytical results, let us
consider the limit of small P\'eclet numbers. For a constant and relatively small force, the P\'eclet function is a constant to quadratic order in the force and we can approximate $\mathrm{Pe}(t)\thickapprox \mathrm{Pe}^{0}$. Denoting it simply as $\mathrm{Pe}$, we can simplify
\be
T_{n}(t)=
\big(n^{2}\,\pi^{2}+\mathrm{Pe^{2}}\big)
\alpha \chi(t),
\ee
so we can readily evaluate
\bea
\!\!\!\sigma^{2}_{L}(t)\!\!&=&\!\!4N_{g}L^{2}\sum_{n=1}^{\infty}
\!\frac{n^{2}\,\pi^{2}\!\big[1\!-\!(-1)^{n}\mbox{e}^{\mathrm{Pe}}\big]^{2}}
{\big[\,n^{2}\,\pi^{2}+\mathrm{Pe}^{2}\big]^{3}}\nonumber\\
&\times&\Bigg[1-\exp\Big[-\frac{n^{2}\,\pi^{2}+\mathrm{Pe}^{2}}{2L^{2}}
\sigma^{2}(t)\,\Big]\Bigg].
\label{MSD-L-Fconst}
\eea
For a bulk fluid with zero force, $\mathrm{Pe}=0$,
\be
T_{n}(t)=
n^{2}\,\pi^{2} \alpha \chi(t),
\ee
and
\bea
\!\!\!\big[\sigma^{2}_{L}(t)\big]_{F=0}\!&=&
\frac{L^{2}}{6}-8L^{2}\sum_{n=1}^{\infty}
\!\frac{1-(-1)^{n}}
{\,n^{4}\,\pi^{4}}\nonumber\\
&\times&\exp\big[-\frac{n^{2}\,\pi^{2}}{L^{2}}
D_{0}\alpha\chi(t)\big].
\label{MSD-L-F0}
\eea
This is consistent with the persistence probability obtained from Eq, (\ref{P(t)-F(t)}) for $\mathrm{Pe}=0$
\be
\!\!\big[P(t)\big]_{_{F=0}}\!
=\!4\!\sum_{n=1}^{\infty}\!
\frac{1\!-\!(-1)^{n}}{n^{2}\,\pi^{2}}
\!\exp\!\Big[\!-\!\frac{n^{2}\,\pi^{2}}{L^{2}}
D_{0}\alpha\chi(t)\Big] .
\label{P(t)-GLE-F0}
\ee
Since the particles in the slab are continuously being absorbed at the boundary, $P(t)$ decays fast to zero and the $\sigma^{2}_{L}(t)$ reaches a plateau $L^{2}/6$. The last is a geometrical factor associated to the phenomenological description inherent to the GLE approach.

To the best of our knowledge, all attempts to study a fluid diffusing within absorbing barriers are based on the Smoluchowski or high friction approximation \cite{redner,szabo2,gitterman,hanggiDybiec,CLOR} which can be obtained as a particular case of our general equations. As $\lambda\rightarrow\infty$ we get the static Langevin approximation, using $\alpha \chi_{_{\mathrm{LE}}}(t)=
(\alpha\,t-1+\mbox{e}^{-\alpha\, t})/\alpha $, while as $\alpha t\gg1$
we get the Smoluchowski limit, using $\alpha \chi_{_{\mathrm{SE}}}(t)= t$ and $D_{z}(t)=D(t)=D_{0}$. In fact, for small constant $\mathrm{Pe}$ numbers, dropping quadratic terms, we let $\mathrm{Pe}=FL/2mD_{0}$, and
we can readily integrate $G(z_{0},t)$ over time to get the following analytical expressions in the SE limiting approximation
\bea
t_{_{\mathrm{MFP}}}^{\mathrm{SE}}(z_{0})
&=&\frac{2L^{2}}{D_{0}}\sum_{n=1}^{\infty}
\frac{n\,\pi\left[1-(-1)^{n}\mbox{e}^{\mathrm{Pe}}\right]}
{[\,n^{2}\,\pi^{2}+\mathrm{Pe^{2}}]^{2}}\nonumber\\
&\times&\sin(n\,\pi \frac{z_{0}}{L}\!)
\exp[-\mathrm{Pe}\frac{z_{0}}{L}],
\label{tMFP-SE}
\eea
\be
\tau^{\mathrm{SE}}
=\frac{4L^{2}N_{g}}{D_{0}}\sum_{n=1}^{\infty}
\!\frac{n^{2}\,\pi^{2}\!\big[1\!-\!(-1)^{n}\mbox{e}^{\mathrm{Pe}}\big]}
{\big[\,n^{2}\,\pi^{2}+\mathrm{Pe}^{2}\big]^{3}},
\label{tau-SE}
\ee
\bea
\!\!\![\sigma^{2}_{L}(t)]^{\mathrm{SE}}\!\!\!\!&=&\!\!\!\!2 D_{0}\tau^{\mathrm{SE}}-4L^{2}N_{g}\sum_{n=1}^{\infty}
\!\frac{n^{2}\,\pi^{2}\!\big[1\!-\!(-1)^{n}\mbox{e}^{\mathrm{Pe}}\big]^{2}}
{\big[\,n^{2}\,\pi^{2}+\mathrm{Pe}^{2}\big]^{3}}\nonumber\\
&\times&\exp\big[-\frac{n^{2}\,\pi^{2}+\mathrm{Pe}^{2}}{L^{2}}
D_{0}t\big],
\label{MSD-L-SE}
\eea

Eq. (\ref{tMFP-SE})
is very close to the expression obtained by Gitterman \cite{gitterman} for the MFPT,
which nevertheless misses the finite exponential terms containing $\mathrm{Pe}$.

For zero force, the sums appearing in Eqs. (\ref{tMFP-SE}) and (\ref{tau-SE}) become the Riemann type, namely, $\sum_{n=1}^{\infty}
 \left[1-(-1)^{n}\right]\sin(n\pi z_{0}/L)/n^{3}=\pi^{3}(L-z_{0})z_{0}/4L^{2}$ and $\sum_{n=1}^{\infty}
\left[1-(-1)^{n}\right]/n^{4}=\pi^{4}/48$, so we get the well known results \cite{hanggiDybiec,hanggiTalknerBorkovec,gitterman}
\bea
\Big[G^{\mathrm{SE}}(z_{0},t)\Big]_{F=0}
&=&2\!\sum_{n=1}^{\infty}
\frac{1-(-1)^{n}}
{\,n\,\pi}\sin\Big(\frac{n\,\pi}{L}
\!z_{0}\Big)\nonumber\\
&\times&\exp\bigg[-\bigg(\frac{n\,\pi}{L}\bigg)^{\!2}
\!D_{0}\,t\bigg],
\label{Gz0t-SE-F0}
\eea
\bea
\big[P^{\mathrm{\mathrm{SE}}}(t)\big]_{F=0}&=&4\sum_{n=1}^{\infty}\!\left[\frac{1-(-1)^{n}}{n\,\pi}\right]\nonumber\\
&\times&\exp\bigg[-\bigg(\frac{n\,\pi}{L}\bigg)^{\!2}
\!D_{0}\,t\bigg],
\label{P(t)SE-F0}
\eea

\be
[t_{_{\mathrm{MFP}}}^{\mathrm{SE}}(z_{0})]_{F=0}= \frac{(L-z_{0})z_{0}}{2D_{0}},
\ee
\be
[\tau^{\mathrm{SE}}]_{_{F=0}} = \frac{L^{2}}{12 D_{0}},
\ee
\bea
\!\![\sigma^{2}_{L}]^{\mathrm{SE}}_{F=0}&=& \frac{L^{2}}{6}
-8L^{2}\sum_{n=1}^{\infty}
\!\frac{1-(-1)^{n}}
{\,n^{4}\,\pi^{4}}\nonumber\\
&\times&\exp\bigg[-\bigg(\frac{n\,\pi}{L}\bigg)^{\!2}
\!D_{0}\,t\bigg].
\eea
\begin{figure}[h]
\includegraphics[width=0.48\textwidth]{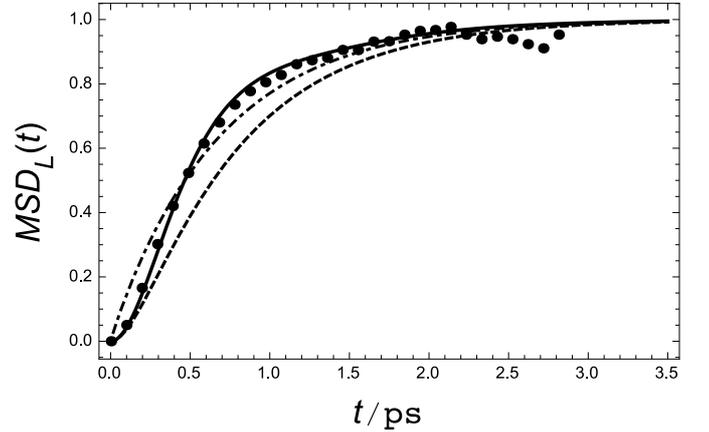}
\caption{The time dependent $ \sigma^{2}_{L}(t)=MSD_{L}(t)$ in an absorbing slab of length $L=1.66\,\mbox{\AA}$, located in the bulk $(a_{_{\mathrm{F}}}=0)$, normalized by its saturation value. GLE (solid line) for $\lambda = 3\,\mathrm{ps}^{-1}$, $\alpha=5.5\,\mathrm{ps}^{-1}$ in Eq. (\ref{MSD-L-F0}), LE (dashed line) and SE (dot--dashed line). The \textcolor{black}{black dots} correspond to the MD simulation\cite{CLOR}.}
\label{FigMSD}
\end{figure}

To test the above equations, we used the MD simulation data for LJ--argon  fully described elsewhere\cite{CLOR}. In Fig. (\ref{FigMSD}) we show the time dependent MSD for particles constrained in an absorbing slab of length $L=1.66\,\mbox{\AA}$, located in the bulk $(a_{_{\mathrm{F}}}=0)$. Values are normalized by its saturation value, namely $MSD_{L}^{*}(t)=\sigma^{2}_{L}(t)/\sigma^{2}_{L}(\infty)$. The theoretical predictions are compared to the MD simulation (black dots)\cite{CLOR}. An almost perfect match was obtained in the GLE (solid line) with an exponential kernel, according to Eqs. (\ref{MSD-L-F0}) and (\ref{chiF}), for $\{\lambda = 3\,\mathrm{ps}^{-1}$, $\alpha=5.5\,\mathrm{ps}^{-1}\}$. The LE (dashed line) approximation predicts the initial ballistic behavior but fails to give the characteristic s shape of the saturation MD curve. The SE (dot--dashed line) approximation predicts only roughly the qualitative behavior.

\section{Conclusions.}
\label{sec:Conclusions}
We have derived analytical formulas for the solution of the non static or generalized Langevin equation (GLE) of a particle in a viscoelastic fluid,
under the action of a time dependent external field.
Emphasis was done in unifying the notation and existing different forms of the master equations, not just for the velocity distribution $P(v,t\mid v_{0},0)$ and its dispersion coefficient $D_{u}(t)$, well discussed in text books\cite{risken,fox1}, but also for the conditional position probability densities $P(z,t\mid z_{0}, v_{0},0)$ and $P(z,t\mid z_{0},0)$ and their corresponding diffusivity coefficients $D_{q}(t)$, $D_{z}(t)$ and the time dependent diffusion coefficient, $D(t)$, associated to the MSD. In doing so, we have extended the results of Dufty\cite{dufty} and Rodriguez {\it et al.} \cite{rodrisalinas} for an unbounded particle to include a time dependent external force. The GLE dynamics of bounded fluid particles constrained by absorbing barriers was then obtained in terms of the results for the unbounded particles.

We wrote all dynamical properties in terms of the fundamental Green functions of the GLE $\chi(t)$, $\chi_{z}(t)$ and $\chi_{v}(t)$, which are the natural
susceptibilities of the system and extended the fluctuation--dissipation theorem to include the external force. Since there is no solution for a general external field, we restricted the equations for the simple constant force case. Thus, we found that for an exponential decaying friction kernel the analytical formulas obtained, like Eq. (\ref{MSD-L-Fconst}), fit well the available exact molecular dynamics simulations. We tested our equations against molecular dynamics data of Argon atoms diffusing between two impenetrable walls separated 40 atomic diameters each other\cite{CLOR}. In particular, at the center of the
system, where the fluid behaves field free, we found that
the mean square displacement in a slab of length 1.66 \AA, normalized to its saturation value, almost
matches the simulation.

We also would like to point out that subsequently from this formal GFPE, we have obtained statistically equivalent sto\-chastic
differential equations for the position of
the non--stat\-ic motion, Eqs. (\ref{newSDE-PJ-GLE}) and (\ref{newSDE-WOR-GLE}), where the noise is white, as in the static Langevin
case, but with time dependent diffusion terms $D_{q}(t)$ and $D_{z}(t)$. The GFPE associated to these alternative non-static SDE have the properties expected for a Brownian
particle. Namely, its stationary solution for the position probability densities $p(z,t|z_{0},v_{0},0)$ and $p(z,t|z_{0},0)$ are Gaussian. In summary, it rigorously treat the Brownian motion in the configuration space, discarding the instantaneous velocity of the particle without resorting to a limit operation of any kind. The new SDE does not impose any restriction on the magnitude of the fluid friction coefficient. For the field free case, it
has the same average and standard deviation
as those prescribed by Chandrase\-khar\cite{chandra}. Moreover, the Smoluchowski limit is guaranteed in
this new approach. This limit was specially useful in problems where the friction term is dominant, such as in
the analysis of the diffusion coefficient in non--homogeneous fluids. They include argon atoms restricted to diffuse
between the force of two impenetrable walls\cite{worpj}, biological systems\cite{szabo} and ionic
fluids\cite{jonsson,ColmenaresPRE}. Our results could be applied to such systems with the interpretation that $D_{z}(t)$ is now a
generalized non--static diffusion coefficient. That is, the existing approaches based on the
Smoluchowski equation would be expanded to the domain of low friction coefficient.

Our GLE describes diffusion that does not obey the normal Brownian
relationships. For instance, the apparent anomalous behavior shown by Eq. (82) is mesoscopically due to a retardation in the friction under a colored noise.

\textcolor{black}{Besides the system studied here, this method of analysis can be perfectly applied to
 exit time problems\cite{gardiner}, controlled diffusion reactions\cite{szabo,redner}, diffusion between reflective barriers\cite{redner}, stochastic thermodynamics\cite{sekimoto} and feed back control in small systems\cite{sagawa}, to mention just a few.
All of them have been analyzed from the perspective of Smoluchowski overdamped motion, so our procedure
can be used to extend them to the demanding underdamped regime. }

\section*{Acknowdledgments}
\textcolor{black}{This work was supported by Universidad de Los Andes through Grant CDCHT-CVI-ADG-C09-95. We thank Floralba L\'opez for optimizing the MD data.}

\renewcommand{\theequation}{A-\arabic{equation}}
  \setcounter{equation}{0}
\section*{Appendix A: The GLE velocity fluctuation coefficients. }
In this Appendix we give a unified view of the existent treatments of the velocity as a stochastic variable. This will led to the evaluation of the fluctuation coefficients associated with the GLE.

  Most of the mathematical treatments for a general stochastic variable or process $\zeta(t)$, have focused on a generalized SDEs of the kind
\bea
 \dot{\zeta}(t)=a(\zeta,t)+ b(\zeta,t)\,\xi(t),
 \label{eqzeta}
\eea
where the drift $a(\zeta,t)$ and the random force intensity $b(\zeta,t)$ are explicit functions of time and the variable $\zeta$.
The existence of formal master equations for this problem has been proven \cite{hanggi}. The result, usually
 obtained via a functional Taylor series of its cumulants, shows that the process $\zeta(t)$ is markovian and
the resulting master equation is Fokker--Planck--like in the Stratonovich sense\cite{gardiner}, {\it i.e.},
\bea
\frac{\partial\, p(\zeta,t)}{\partial t} &=&-\frac{\partial}{\partial \zeta}\left[a(\zeta,t)\,p(\zeta,t)\right]\nonumber\\
&+&\frac{1}{2}\,\frac{\partial}{\partial \zeta}\left[b(\zeta,t)\,\frac{\partial}{\partial \zeta}b(\zeta,t)\,p(\zeta,t)\right].
\label{FPEhanggi}
\eea
But we should point out that, using the relationship between the Stratonovich drift velocity first moment $\langle \dot{\zeta}\rangle_{_{\!\mathrm{S}}}$ and Ito first moment $\langle \dot{\zeta}\rangle_{_{\!\mathrm{I}}}$
\be
\langle \dot{\zeta}\rangle_{_{\!\mathrm{S}}}=\langle \dot{\zeta}\rangle_{_{\!\mathrm{I}}}+\frac{1}{2}\,\frac{\partial}{\partial \zeta}b^{2}(\zeta,t),
\ee
the corresponding FPE in the Ito sense is
\bea
\frac{\partial\, p(\zeta,t)}{\partial t} &=&-\frac{\partial}{\partial \zeta}\left[a(\zeta,t)\,p(\zeta,t)\right]\nonumber\\
&+&\frac{1}{2}\,\frac{\partial^{2}}{\partial \zeta^{2}}\left[b^{2}(\zeta,t)\,p(\zeta,t)\right].
\label{FPEhanggiITO}
\eea
When $\zeta(t)$ is the velocity $v(t)$, Eq. (\ref{FPEhanggiITO}) is just a Rayleigh-type equation for the velocity probability $p(v,t\mid v_{0})$, given an initial velocity $v_{0}$. In fact, for $a(v,t)=a(t)=\overline{\dot{v}(t)}$
  and $b(v,t)=b(t)=\eta(t)/\gamma(t)=\sqrt{2D_{u}(t)}$, independent of $v(t)$, we get the master equation
  associated to the LE, Eq. (\ref{setLE}) in the absence of an external field, i.e. $F(z,t)=0$, namely
\bea
\Bigg(\frac{\partial p(v,t\mid v_{0})}{\partial t}\Bigg)_{\!\!v}&=&-\dot{\overline{v}}(t)\,\frac{\partial \,p(v,t\mid v_{0})}{\partial v}\nonumber\\
&+&D_{u}(t)\,\frac{\partial^{2}p(v,t\mid v_{0})}{\partial v^{2}},
\label{FPE-Rayleigh}
\eea
where $\overline{v}(t)\!=\!\langle v(t) \rangle_{\xi}$ is the average of $v(t)$ over the noise $\xi$. We have used the
subscript $u$, instead of $v$, in the fluctuation coefficient $D_{u}$, anticipating that the resulting Eq. (\ref{FPE-Rayleigh})
has the standard form of the Kramers-Moyal master equation expansion, with a first drift parameter in terms of the rate of the first
moment $\overline{v}$ and a second dispersion parameter in terms of the rate of the second moment of
$u(t)=\big[v(t)-\overline{v}\big]$, namely, $2D_{u}(t)=d\,\overline{u^{2}(t)}/dt$. The appropriate application of Eq. (\ref{FPEhanggiITO}) when the stochastic variable $\zeta(t)$ is
set as the particle position $z(t)$, is discussed in the text.

Using the standard Chandrasekhar's arguments\cite{chandra,mcquarrie}, the quantity $u(t)$ must have the same
 Gaussian statistics as the noise, so the velocity conditional probability density which satisfies Eq. (\ref{FPE-Rayleigh}) can be readily written as
\be
p(v,t\mid v_{0})\!=\! \frac{1}{\sqrt{2\,\pi\,\sigma_{\!u}^{2}(t)}}\exp\left[-\frac{\big[v-\overline{v}(t)\big]^{2}}{2\,\sigma_{\!u}^{2}(t)}\right],
\label{chandravelocity}
\ee
where $d\,\sigma_{\!u}^{2}(t)/dt=2\,D_{u}(t)$. Writing the Gaussian in the alternative form
 \be
 \Bigg(\frac{\partial\, p(v,t\mid v_{0})}{\partial v(t)} \Bigg)_{\!\!t} =-\frac{\big[v(t)-\overline{v}(t)\big]}
 {\sigma_{\!u}^{2}(t)}\,p(v,t\mid v_{0}),
 \ee
 we can rewrite Eq. (\ref{FPE-Rayleigh}) in Adelman's Fokker-Planck form \cite{adelman1}
\bea
\Bigg(\frac{\partial\, p(v,t\mid v_{0})}{\partial t}\Bigg)_{\!\!v}&=&\beta(t)\,\frac{\partial \big[v(t)\,p(v,t\mid v_{0})\big]}{\partial v}\nonumber\\
&+& D_{v}^{\mathrm{A}}(t)\,\frac{\partial^{2}p(v,t\mid v_{0})}{\partial v^{2}},
\label{ME-Adelman}
\eea
 where $\beta(t)=-\dot{\overline{v}}(t)/\overline{v}(t)$ and the apparent coefficient $D_{v}^{\mathrm{A}}(t)$ is
 \be
D_{v}^{\mathrm{A}}(t)=D_{u}(t) + 2\,\beta(t)\int_{0}^{t}D_{u}(s)ds.
\label{DFP}
 \ee
 This form of the master equation for the generalized GLE, Eq.(\ref{GLE}), is
 commonly referred to as the velocity space Fokker-Planck equation ($v$-FPE)\cite{adelman1,fox1,hanggiTalkner}.
  Next, we evaluate the properties of the colored noise functions and the jump moments corresponding to the
probability densities $p(v,t\!\!\mid \!\!v_0) $ and $p(v,t)$ for the velocities in Chandrasekhar's context\cite{chandra}.

According to Eqs. (\ref{vGLE}) and (\ref{varphi}), for a Gaussian color noise $R(t)$, the definition of the auxiliary velocity function
\be
u(t)=v(t)-\overline{v}(t)=\varphi_{v}(t),
\label{App-u(t)}
\ee
with $\overline{v}(t)=v_{0}\chi_{v}(t) + \phi_{v}(t)$, ensures,
a diffusion-like equation for the velocity conditional probability density $ p(u,t\!\!\mid \!\!v_{0}) $
\be
\Big(\frac{\partial\, p(u,t\!\!\mid \!\!v_{0})}{\partial t}\Big)_{u(t)}=D_{u}(t)\,\frac{\partial^{2}\,p(u,t\!\!\mid \!\!v_{0})}{\partial u^{2}}.
\label{v-Dequation}
\ee

 Equations (\ref{FPEhanggiITO}), (\ref{FPE-Rayleigh}), (\ref{ME-Adelman}),
  and (\ref{v-Dequation}) are all equivalent master equations in $v$-space for the GLE $p(v,t\mid v_{0})$, when
 the proper definitions of the diffusion coefficients $D_{u}(t)$ and $D_{u}^{\mathrm{A}}(t)$ are used.
The solution of Eq. (\ref{v-Dequation}) in an unbounded space is the standard Gaussian
\be
 p(u,t\!\!\mid \!\!v_{0})\!=\! \frac{1}{\sqrt{2\,\pi\,\sigma_{\!u}^{2}(t)}}\exp\left[-\frac{u^{2}(t)}{2\,\sigma_{\!u}^{2}(t)}\right].
\label{App-chandraprob}
\ee

The first moment is simply $\overline{u}(t)=\langle u(t)\rangle_{_{\!R}} = 0$
and $\langle v(t)\rangle_{_{\!R}} = \overline{v}(t)=v_{0}\chi_{v}(t)+\phi_{v}(t)$. The velocity drift $\phi_{v}(t)$ is defined in Eq. (\ref{varphiF}) and reduces to $a_{_{\mathrm{F}}}\chi_{z}(t)$ for a constant force.

The second moment, $\sigma_{u}^{2}(t)=\langle u^{2}(t)\rangle_{_{\!R}}$, can be obtained from Eq. (\ref{App-u(t)})
\be
\sigma_{u}^{2}(t)=\langle u^{2}(t)\rangle_{_{\!R}}=\langle \varphi_{v}^{2}(t)\rangle_{_{\!R}}.
\label{App-1}
\ee

The first identity, together with the solution of the GLE, Eqs. (\ref{vGLE}) and (\ref{vtilde}), gives
\be
\sigma_{u}^{2}(t)=\langle v^{2}(t)\rangle_{_{\!R}} -v_{0}^{2}\chi_{v}^{2}(t) -\phi_{v}^{2}(t)
-2v_{0}\phi_{v}(t).
\label{App-1a}
\ee

Since $\sigma_{u}^{2}(t)$ is independent of the initial conditions, one can average this identity over $v_{0}$ and
use the physical stationary and equipartition conditions
\be
\big\langle \langle v^{2}(t)\rangle_{_{\!R}} \big\rangle_{\!v_{0}}=
\langle v_{0}^{2}\rangle_{_{\!v_{0}}}
=\frac{k_{_{\mathrm{B}}}T}{m},
\label{App-conditions}\ee
with $\langle v_{0}\rangle_{_{\!v_{0}}}=0$, in order to get
\be
\sigma_{u}^{2}(t)=\frac{k_{_{\mathrm{B}}}T}{m}\,[ 1 - \chi_{v}^{2}(t)] -\phi_{v}^{2}(t),
\label{App-Sigmau}\ee
and therefore the diffusion like velocity coefficient $D_{u}(t)$ in Eq. (\ref{v-Dequation}) is
$ D_{u}(t)=(1/2)d\,\sigma_{u}^{2}(t)/dt$
\bea
D_{u}(t)&=&\frac{k_{_{\mathrm{B}}}T}{m} \beta(t)\,\chi^{2}_{v}(t) \!-\!\phi_{v}(t)\dot{\phi}_{v}(t),
\label{App-Du}
\eea
where
\be \beta(t)=-\frac{\dot {\chi}_{v}(t)}{\chi_{v}(t)}\label{App-beta},
\ee

\begin{center}
\begin{figure}[ht]
\includegraphics[width=0.48\textwidth]{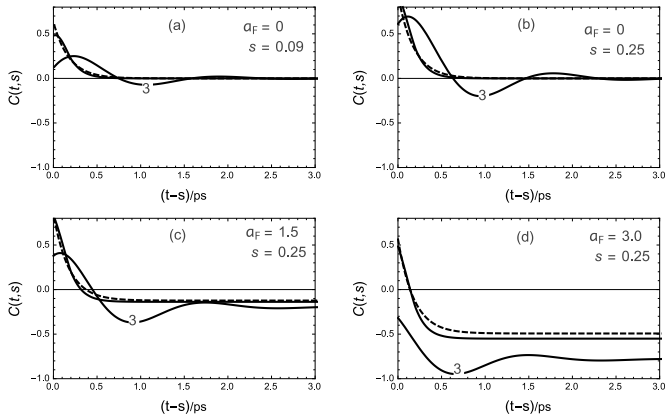}
\caption{The free diffusion two--time correlation function
$C(t,s)=\langle\overline{\varphi_{_{v}}(t)
\varphi_{_{v}}(s)}\rangle $ vs $(t-s)$, scaled by $(D_{_{0}}\,\alpha)$. Graphs (a) and (b) are for a null external force at two initial times: (a) a low $s=0.09\, \mbox{ps}$ and (b) a higher $s=0.25\, \mbox{ps}$. Graphs (b), (c) and (d) show the effect of a constant external force: $a_{_{\mathrm{F}}}$ in $\mbox{\AA}/\mathrm{ps}^{2}$ of 0, 1.5 and 3.0 respectively. In all graphs a low and a high values of $\lambda$ were used for GLE (solid lines), $3\, \mbox{ps}^{-1}$ (labeled) and $25\, \mbox{ps}^{-1}$. Dashed curves correspond to LE. $\alpha=5.5\, \mbox{ps}^{-1}$.}
\label{Fig3}
\end{figure}
\end{center}

 These results can be obtained also from the second identity of Eq. (\ref{App-1}) involving the color noise force function $\varphi_{v}(t)$ defined in Eq. (\ref{varphi})\cite{dufty,adelman1}
\be
\sigma_{u}^{2}(t)\!\!=
\,\!\!\!\!\int_{0}^{t}\!\!\!\chi_{v}(t\!-\!t^{\prime})\,dt^{\prime}\!\!\!
\int_{0}^{t}\!\!\!\chi_{v}(t\!-\!t^{\prime\prime})
\big\langle R(t^{\prime})R(t^{\prime\prime})\,\big\rangle_{_{\!\!R}}\!dt^{\prime\prime}.
\label{App-sigmavarphi}\ee

The Gaussian R-noise two time correlation appearing in this expression must satisfy a physical consistency relation, which is widely known as the fluctuation-dissipation theorem (FDT). For a time dependent external force $F(t)$
\bea
\frac{1}{m^{2}}\Big(\big\langle R(t^{\prime})R(t^{\prime\prime})\big\rangle_{_{\!R}}&+& F(t^{\prime})F(t^{\prime\prime})\Big)=\nonumber\\
&&\frac{k_{_{\mathrm{B}}}T}{m}\,\Gamma(\mid t^{\prime}-t^{\prime\prime} \mid) ,\label{App-FDTheorem}
\eea
\bea
\Big\langle R(t^{\prime})R(t^{\prime\prime})\Big\rangle_{_{\!R}}&=&k_{_{\mathrm{B}}}Tm\,\Gamma(\mid t^{\prime}-t^{\prime\prime} \mid)\nonumber\\
 &-& F(t^{\prime})F(t^{\prime\prime}).
\eea
 This relationship is derived from Eqs. (\ref{GLE}) and (\ref{chivdot})\cite{BerneHarp}, through the correlation of the dissipative force
 $\Gamma(\mid t^{\prime}-t^{\prime\prime}\mid)$ with both the internal and external fluctuation force $R(t)+F(t)$, assuming the stationary and equipartition conditions, therefore it is equivalent to Eq.(\ref{App-conditions}). Here we have an extra drift term due to the external force which is often neglected in the literature \cite{balak,cicotti}. This theorem is useful for evaluating the two--time correlation function $C(t,s)=\Big\langle\Big\langle\varphi_{v}(t)\,\varphi_{v}(s)\Big\rangle_{R}\Big\rangle_{v_{0}}$; it is given by:
\bea
C(t,s)&=&\frac{1}{m^{2}}\int_{0}^{t}\chi_{v}(t-t^{\prime})\,dt^{\prime}\nonumber\\
&\times&\int_{0}^{s}\chi_{v}(s-t^{\prime\prime})\,\Big\langle\overline{R(t^{\prime})R(t^{\prime\prime})}\Big\rangle \,dt^{\prime\prime}.
\label{corrGLE1}
\eea
It was nicely shown by Adelman\cite{adelman1} and Fox\cite{fox1} that
\bea
\int_{0}^{t}\!\!\chi_{v}(t-t^{\prime})\,dt^{\prime}
\int_{0}^{s}\!\!\chi_{v}(s-t^{\prime\prime})\,
\Gamma(\mid t^{\prime}-t^{\prime\prime}\mid)\,dt^{\prime\prime}\nonumber\\
=\chi_{v}(\mid t-s \mid)-\chi_{v}(t)\chi_{v}(s).
\label{App-chiFox}
\eea
Using this, together with Eq. (\ref{App-FDTheorem}), in Eq.(\ref{App-sigmavarphi}) one obtains the general result
\bea
C(t,s)&=&\frac{k_{_{\mathrm{B}}}T}{m}\big[\chi_{v}(\mid t-s \mid)-\chi_{v}(t)\chi_{v}(s)\big]\nonumber\\
&-&\phi_{v}(t)\phi_{v}(s).
\label{corfox}
\eea
Using $t=s$ and $\chi_{v}(0)=1$, Eq. (\ref{corfox}) reduces to Eq. (\ref{App-Sigmau}) as expected.

In Fig. (\ref{Fig3}), the two time correlation function $C(t,s)$ is shown as a function of the time difference $(t-s)$, for two initial times $s=0.25\,\mbox{ps}$ (top) and $1.5\,\mbox{ps}$ (bottom) and two values of the frequency $\lambda$ as labeled. Static Langevin theory (LE) gives a single exponential decay (dotted curve), as derived in Appendix B, Eq. (\ref{corrLE}). For $\lambda$ as large as $25\,\mbox{ps}^{-1}$, GLE already approaches the LE. Since the frequency $\omega$ was defined as a real positive quantity for $4\,\alpha>\lambda$, the susceptibility function $\chi_{v}(t)$ is an oscillatory function of time for low values of $\lambda$. Hence a negative correlation is expected from the definition of $C(t,s)$, Eq. (\ref{corfox}), particularly for the higher values of the external force.

 \renewcommand{\theequation}{B-\arabic{equation}}
  \setcounter{equation}{0}
\section*{Appendix B: The static-LE and high friction-SE limits.}
\label{sec:Static}
In this section, we establish the connection of our general results with the commonly used approximations in the static and high friction limits. To find the drift and diffusion terms in the static LE for a constant external force, we apply the procedure of Sections (\ref{sec:I Generalized}) and (\ref{sec:III Analytical}) to Eq. (\ref{setLE}). The result is similar to Eq.(\ref{z-GRFPE}), with $\lambda\rightarrow \infty$. The characteristic functions become
$\chi_{v}^{\mathrm{LE}}=e^{-\alpha t}$, $\alpha
\chi_{z}^{\mathrm{LE}}=1-e^{-\alpha t}$, and
$\alpha^{2}\chi^{\mathrm{LE}}= e^{-\alpha t} + \alpha t - 1$. Furthermore,
\bea
\overline{v}_{_{\mathrm{LE}}}(t) = v_{_{0}}\mbox{e}^{-\alpha t} + v_{_{\mathrm{F}}}(1-\mbox{e}^{-\alpha\,t}),\label{driftLE}
\eea
\bea
[\sigma_{u}^{2}(t)]_{_{\mathrm{LE}}} = D_{0}\,\alpha\,(1-\mbox{e}^{-2\,\alpha\, t})-v_{_{\mathrm{F}}}^{2}(1-e^{-\alpha t})^{2},
\eea
\bea
[D_{u}(t)]_{_{\mathrm{LE}}} = D_{0}\,\alpha^{2}\,\mbox{e}^{-2\,\alpha\, t} -v_{_{\mathrm{F}}}^{2} \alpha (1-e^{-\alpha t})e^{-\alpha t}.
\label{app-DvLE}
\eea
The moments associated to the probability density \,\,\,
 $p(z,t\mid z_{0},v_{0})$ are
\bea
\langle\overline{z}(t)\rangle^{^{\mathrm{LE}}}&=&
z_{_{0}}+\frac{v_{_{\!\mathrm{0}}}}{\alpha}\big(1-\mbox{e}^{-\alpha t}\big)\nonumber\\
&+&\frac{v_{_{\!\mathrm{F}}}}{\alpha}\big(\mbox{e}^{-\alpha t}+\alpha t -1\big),
\eea
\bea
[\sigma^{2}_{q}(t)]_{_{\mathrm{LE}}} &=&
\frac{D_{_{0}}}{\alpha}\left[2\,\alpha t-3+4\,\mbox{e}^{-\alpha\, t}-\mbox{e}^{-2\,\alpha\,t}\right]\nonumber\\
&-&\frac{v^{2}_{_{\mathrm{F}}}}{\alpha^{2}}
\big(\mbox{e}^{-\alpha t}+\alpha t -1\big)^{2}.
\label{sigqLE}
\eea
For $v_{_{\mathrm{F}}}=0$ this reduces to the textbook result\cite{mcquarrie}.
The $v_{0}$-average probability $p_{_{_{\mathrm{\!LE}}}}\!(z,t|z_{_{0}})$ given by Eq. (\ref{Probfinal}), evolves
in time as an equation similar to Eq.(\ref{z-GFPE}) with fluctuation moments:
\be
\langle\overline{z}(t)\rangle^{^{\mathrm{LE}}}_{_{\mathrm{v_{0}}}}
=z_{_{0}} +\frac{v_{_{\!\mathrm{F}}}}{\alpha}\big(\mbox{e}^{-\alpha t}+\alpha t -1\big)
\ee
\bea
[\sigma^{2}_{z}(t)]_{_{\mathrm{LE}}}
&=&\frac{2D_{_{0}}}{\alpha}\big(\mbox{e}^{-\alpha t}+\alpha t -1\big) \nonumber\\
&-&\frac{v^{2}_{_{\mathrm{F}}}}{\alpha^{2}}
\big(\mbox{e}^{-\alpha t}+\alpha t -1\big)^{2},
\label{PhiLE}
\eea
\bea
[D_{z}(t)]_{_{\mathrm{LE}}}&=&D_{_{0}}\big( 1-\mbox{e}^{-\alpha\, t}\big)\nonumber\\
&\times&\big[ 1 -\frac{v_{_{\mathrm{F}}}^{2}}{D_{_{0}}\alpha}
\big(\mbox{e}^{-\alpha t}+\alpha t -1\big)\big],
\label{app-DzLE}
\eea
The plain $\textsl{MSD}(t)$ and diffusivity coefficient $D(t)$ are
\be
[\textsl{MSD}(t)]_{_{\mathrm{LE}}} =
\frac{2D_{_{0}}}{\alpha}\big( \mbox{e}^{-\alpha t}+\alpha t -1 \big),
\ee
\be
[D(t)]_{_{\mathrm{LE}}} =
D_{0} \big( 1-\mbox{e}^{-\alpha\, t}\big).
\label{D(t)-LE}
\ee
The apparent diffusivity $\widetilde{D}(t)$ is
\be
[\widetilde{D}(t)]_{_{\mathrm{LE}}}=
\frac{D_{_{0}}}{\alpha t}\big(\mbox{e}^{-\alpha t}+\alpha t -1 \big),
\ee
corresponding to the colored noise $\varphi_{_{\mathrm{LE}}}(t)$,
whose two--time correlation function for $t>s$ is just the limit of Eq. (\ref{corfox}) for $\lambda\rightarrow\infty$
\bea
\Big\langle\varphi_{v}(t)
\varphi_{v}(s)\Big\rangle^{_{\mathrm{LE}}}&=&
D_{_{0}}\alpha\Big(\mbox{e}^{-\alpha\,\mid t-s\mid}-\mbox{e}^{-\alpha\,(t+s)}\Big)\nonumber\\
&-&v_{_{F}}^{2}(1-e^{-\alpha t})(1-e^{-\alpha s}).
\label{corrLE}
\eea
All of this agrees with the result of applying Eq. (\ref{varphi}) consistently, with
 $\Gamma_{_{\mathrm{LE}}}(t)=\alpha\,\,\delta(t)$ and
\be
\frac{1}{m^{2}}\langle
R(t)R(s)\rangle_{_{\mathrm{LE}}}=\frac{k_{_{\mathrm{B}}}T}{m}\, \alpha\,\delta(t-s)- v_{_{\mathrm{F}}}^{2}\alpha^{2},
\ee
 respectively.
We notice that both, the LE and its generalization GLE, have a Gaussian probability distribution. In particular,
for the static LE, Eq. (\ref{chandra}) with zero external force, i.e. $v_{_{\mathrm{F}}}=0$,
matches Eq. (171) of Chandrasekhar paper on Brownian motion of a 1--D free particle\cite{chandra}. Therefore, at
large times $\sigma^{2}_{_{\mathrm{LE}}}(t)\rightarrow 2D_{_{0}}t$.
 The so called high friction limit (HFL) corresponds to the limit $\alpha\rightarrow\infty$. Physically, it is consistent
 with a small relaxation time $1/\lambda$ of the friction retardation kernel. So, the HFL of the GLE requires that we take first the static LE limit, $\lambda\rightarrow\infty$, and $\alpha\rightarrow\infty$, afterwards. Since, the first limit
corresponds to the static LE, then from Eqs. (\ref{driftLE}) and (\ref{app-DzLE}), the HFL gives
$\overline{v}_{_{\mathrm{HFL}}}(t)=v_{_{\!\mathrm{F}}}$ and ${D_{_{\mathrm{\!HFL}}}(t)=D_{_{0}}}$. Therefore,
 Eq. (\ref{newSDE-PJ-GLE}) becomes under this condition
 \be
 \frac{dz(t)}{dt}=v_{_{\!F}}+\sqrt{2\,D_{_{0}}}\,\xi(t).
 \label{IRA}
 \ee
 Likewise, Eq. (\ref{z-GRFPE}) reduces to the so called Smoluchowski equation (SE). The probability density
 $p_{_{\mathrm{HFL}}}(z,t)$ is a Gaussian with the moments
$\overline{z}(t)_{_{\mathrm{\!HFL}}} = z_{_{0}}+v_{_{\!\mathrm{F}}} t$ and
$\sigma^{2}_{_{\mathrm{\!HFL}}}(t)=2 \,D_{_{0}}\,t$.
Another way to get Eq. (\ref{IRA}) is by taking first the limit $\lambda\to\infty$ to Eq. (\ref{corrGLE1}) and then
the limit of large $\alpha$. As expected, the first result reproduces Eq. (\ref{corrLE}), and then for large friction, the correlation $\langle\varphi(t)\varphi(s)\rangle$
is a Dirac delta function and the noise $\varphi(t)$ transforms into white noise in this limit\cite{gardiner}. Similarly, the
drift $\overline{v}_{_{\mathrm{LE}}}(t)$ reduces to $v_{_{\!\mathrm{F}}}$ and Eq. (\ref{newSDE-PJ-GLE}) becomes Eq. (\ref{IRA}).

\section*{References}

\end{document}